\documentclass[aps,pre,twocolumn,groupedaddress,showpacs,superscriptaddress,amssymb,amsmath]{revtex4-2}
\usepackage[utf8]{inputenc}
\usepackage[T1]{fontenc}
\usepackage[english]{babel}
\usepackage{graphicx}
\usepackage{epsfig}
\usepackage{epstopdf}
\usepackage{amsmath}
\usepackage{amssymb}
\usepackage{times}
\usepackage{setspace}
\usepackage{verbatim}
\usepackage{color}
\usepackage{mathrsfs}
\usepackage{euscript}
\usepackage{mathptmx}

\usepackage{hyperref}
\hypersetup{
    colorlinks=true,
    linkcolor=blue,
    filecolor=magenta,      
    urlcolor=blue,
    citecolor=blue, 
    }

\usepackage[normalem]{ulem}

\begin{document}
\title{Finite temperature equilibrium density profiles of integrable systems in confining potentials}

\author{Jitendra Kethepalli}
\email{jitendra.kethepalli@icts.res.in}
\address{International Centre for Theoretical Sciences, Tata Institute of Fundamental Research, Bengaluru -- 560089, India}

\author{Debarshee Bagchi}
\email{debarshee.bagchi@icts.res.in}
\address{International Centre for Theoretical Sciences, Tata Institute of Fundamental Research, Bengaluru -- 560089, India}

\author{ Abhishek Dhar}
\email{abhishek.dhar@icts.res.in}
\address{International Centre for Theoretical Sciences, Tata Institute of Fundamental Research, Bengaluru -- 560089, India}

\author{Manas Kulkarni}
\email{manas.kulkarni@icts.res.in}
\address{International Centre for Theoretical Sciences, Tata Institute of Fundamental Research, Bengaluru -- 560089, India}

\author{Anupam Kundu}
\email{anupam.kundu@icts.res.in}
\address{International Centre for Theoretical Sciences, Tata Institute of Fundamental Research, Bengaluru -- 560089, India}

\date{\today}

\begin{abstract}
We study the equilibrium density profile of particles in two one-dimensional classical integrable models, namely  hard rods and the hyperbolic Calogero model, placed in confining potentials. For both of these models the inter-particle repulsion is strong enough to prevent particle trajectories from intersecting. We use field theoretic techniques to compute the density profile and their scaling with system size and temperature, and compare them with results from Monte-Carlo simulations. In both cases we find good agreement between the field theory and simulations. We also consider the case of the Toda model in which inter-particle repulsion is weak and particle trajectories can cross. In this case, we find that a field theoretic description is ill-suited due to the lack of a thermodynamic length scale. The density profiles for the Toda model obtained from Monte-Carlo simulations can be understood by studying the analytically tractable harmonic chain model (Hessian approximation of the Toda model). For the harmonic chain model one can derive an exact expression for the density that shines light on some of the qualitative features of the Toda model in a quadratic trap. Our work provides an analytical approach towards understanding the equilibrium properties for interacting integrable systems in confining traps.
\end{abstract}

\maketitle
\section{Introduction}
\label{intro}
Integrable classical systems \cite{babelon2003introduction,torrielli2016classical,cao1995classical} have a macroscopic number of constants of motion that are in involution with each other. In phase space, these systems (i) have regular periodic orbits (invariant torus), (ii) are characterized by zero Lyapunov exponents, and (iii) generally resist thermalization to a Gibbs state. However, many-body integrable systems are also believed to be extremely fragile in the presence of external perturbations, and become nonintegrable, ergodic, and chaotic, retaining only a few constants of motion \cite{bastianello2021hydrodynamics}. Consequently, integrable systems are rare, and nonintegrability arising due to imperfections dominates the natural world. For example, most experiments \cite{kinoshita2006quantum,tang2018thermalization,bouchoule2022generalized} are performed in confining potentials where we expect that integrability will be lost and thermalization to occur. Recent theoretical studies have addressed thermalization and transport in such trapped integrable models \cite{di2018transport,lam2014stochastic,rajabpour2014quantum,cao2018incomplete,dhar2019transport,caux2019hydrodynamics,durnin2021nonequilibrium,bulchandani2021quasiparticle}.
To study thermalization, one needs to have a clear understanding of the thermal equilibrium state. One simple characterization is to look at the equilibrium density profile of the particles in the trap which is the most commonly measured quantity in experiments \cite{inguscio2008ultra,dalfovo1999theory,joseph2011observation}.

In this work, we focus on equilibrium density profile of two one-dimensional short-range integrable classical models in the presence of integrability-breaking external potentials. The integrable models considered here are the gas of hard rods  \cite{tonks1936complete} and the hyperbolic Calogero model \cite{olshanetsky1981classical,perelomov1990integrable,polychronakos1992new}. The external trap potential keeps the particles spatially confined and breaks integrability. Such systems have been studied recently and many surprising results have been reported. For example the gas of hard rods~\cite{cao2018incomplete} and the Lieb-Liniger model~\cite{caux2019hydrodynamics} in quadratic trap were investigated. It was found that these systems do not thermalize even in the presence of the quadratic trap. Under out-of-equilibrium conditions, drastically slow relaxations to a nonequilibrium steady state and large finite size effects have also been observed for the Toda model with harmonic (quadratic) pinning potential \cite{di2018transport,dhar2019transport}.
In another recent work \cite{fu2019universal}, a similar observation was made for the nonlinearly perturbed Toda model and a universal scaling of the thermalization time has been reported.
Studies of the integrable Calogero model in the presence of external confining potentials have also been undertaken in recent times, see for example Refs.  \cite{gurappa1999equivalence,rajabpour2014quantum,kulkarni2017emergence,gon2019duality,bulchandani2021quasiparticle}.

The equilibrium properties of trapped interacting particles have recently been studied where field theoretic techniques are used to compute the equilibrium density profiles and fluctuations \cite{dean2006large,majumdar2014top,agarwal2019harmonically,kumar2020particles,kethepalli2021harmonically,kethepalli2022edge,santra2022gap}. Here we adapt these field theoretic procedures to study the equilibrium properties of hard rods and the hyperbolic Calogero system in the presence of external trapping potentials.
The field theory presented here predicts quite accurately the equilibrium density profile of these two models, and their scaling with system size and temperature, as obtained from Monte-Carlo (MC) simulations.

However, we find that the behavior of trapped integrable models with weak repulsion, such as the Toda model, appears to be strikingly different from the above-mentioned models. It turns out that for such short-range models where particle crossings are allowed, the density profile is localized on a length scale that is system size independent, thereby rendering the field theoretic description ill-suited. In a suitable parameter regime, the equilibrium density profiles of the Toda model can be understood using the Hessian approximation of the Toda interaction which is the nearest neighbor harmonic chain. Note that the harmonic chain model in the quadratic trap is analytically tractable and thereby provides a transparent way of understanding the density profiles.

The paper is organized as follows. We describe the models and definitions in Sec.~\ref{models}. Thereafter, in Sec.~\ref{ftf}, we present the field theory for the hard rods gas and the hyperbolic Calogero model in both quadratic and quartic traps. In Sec.~\ref{density} we compute the densities and extract their scaling with system size and temperature for (i) hard rods gas in Sec.~\ref{hrd} and (ii) hyperbolic Calogero model in Sec.~\ref{hcm}. We verify the analytical results using Monte-Carlo (MC) simulations. Next, in Sec.~\ref{int}, we study the equilibrium density profiles of the Toda model along with its Hessian approximation in a suitable parameter regime. We summarize the main results in Sec.~\ref{conclusion} and end with a discussion of open questions in such integrability broken classical systems. The Appendix is organised as follows. In Appendix~\ref{AppI}, we derive the field theory for (i) hard rods gas and (ii) hyperbolic Calogero model in external confining traps. In Appendix~\ref{AppII}, we compute the analytical form of the densities for low and high values of the temperature. In Appendix~\ref{AppIII}, we derive the equilibrium density profile for the quadratically confined nearest neighbour harmonic chain.
\section{Models and definitions}
\label{models}
We study two short-range models given by a Hamiltonian of the form
\begin{equation}
H(\{x_i, p_i\}) = \sum_{i=1}^N \left[\frac {p_i^2}{2m} + U_{\delta}(x_i)\right] +  \frac{1}{2}\sum_{i=1}^N \sum_{\substack{j=1 \\ j\neq i}}^{N} V(x_i - x_j),
\label{H}
\end{equation}
where $\{x_i, p_i\}$ are the position and momentum of the $i^{\rm th}$ particle ($1 \leq i \leq N$), each of mass $m$ which we set to unity. The second term on the right-hand side of Eq. (\ref{H}) is the external potential 
\begin{align}\label{external}
	U_{\delta}(x) = \frac{x^\delta}{\delta},
\end{align}
which we take to be of quadratic ($\delta = 2$) or quartic ($\delta = 4$) form. The third term in Eq. (\ref{H}) is the interaction term, which for hard rods (HR) of length $a$ is 
\begin{equation}\label{hrd:inter}
V_R(r) = 
\begin{cases} 
0 & \text{~for~} \quad r > a \\
\infty & \text{~for~} \quad r \leq a.
\end{cases}
\end{equation}
Note that in Eq.~\eqref{hrd:inter} the subscript `$R$' in $V_R(r)$ stands for the hard rods gas.
For the hyperbolic Calogero (HC) model each particle is coupled to every other particles in the system with the interaction potential
\begin{equation}\label{hcm:inter}
V_C(r) = \frac{J}{\sinh^2|r|}.
\end{equation}
In Eq.~\eqref{hcm:inter}, the subscripts `$C$' in $V_C(r)$ stand for the hyperbolic Calogero model and $J>0$ is the strength of the repulsive interaction. 

We consider these systems to be in their respective thermal equilibrium states described by the canonical Gibbs distribution 
\begin{equation}
P(\{x_i, p_i\}) = \frac {e^{-\beta H(\{x_i, p_i\})}}{Z_{\beta}(N)},
\label{P}
\end{equation}
where $\beta = 1/T$ is the inverse temperature and $Z_{\beta}(N)$ is the partition funtion.
We are interested in the spatial density profile
\begin{equation}
\rho(x) = \sum_{i=1}^N \left< \delta(x-x_i)\right>_{\beta},
\end{equation}
where $\langle \ldots \rangle_{\beta}$ denotes the average over the thermal distribution given in Eq.~\eqref{P}. In particular, we will examine the dependence of the density profile $\rho(x)$ on system parameters, such as the number of particles $N$ and the temperature $T$. In the following sections, we address these questions using field theory and MC simulations.

\begin{figure}[htb]
	\centering
	{\includegraphics[width=8cm]{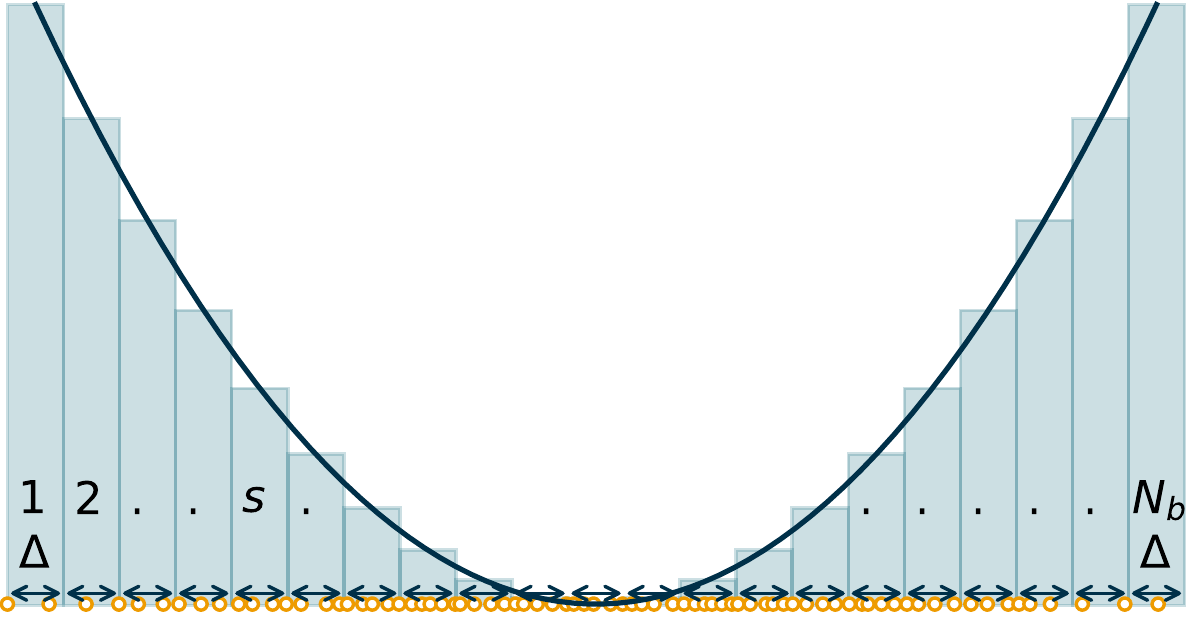}}
	\caption{A schematic representation for partitioning the system into $N_b$ subsystems, each of size $\Delta$. Here $s = 1,2,\ldots,N_b$ denotes the subsystem index. Note that the particles are represented by orange circles. In the $s^{\rm th}$ subsystem there are $n_s$ particles. The double arrow indicates the extent of each subsystem. While analyzing each subsystems we take large-$n_s$ and then for the complete system we finally take small-$\Delta$ limit.}
	\label{fig:schematic}
\end{figure}

\section{Field theory formalism}
\label{ftf}
To obtain the thermal properties one needs to compute the partition function $Z_{\beta}(N)$ which is generally a hard task in microscopic variables. Therefore often one resorts to field theoretic (macroscopic) approach to compute $Z_{\beta}(N)$. In this method, the partition function is written as a functional integral over density fields. This procedure has been commonly used in several contexts such as Landau theory~\cite{mussardo2010statistical}, random matrix theory ~\cite{dean2006large}, general Coulomb gas~\cite{cunden2018universality}, and long-range interacting particles ~\cite{hardin2018large, agarwal2019harmonically}. Despite this progress there has been only a few rigorous comparisons between densities and other equilibrium properties obtained from microscopic and macroscopic (field theory) computation~\cite{allez2012invariant,agarwal2019harmonically, kumar2020particles,kethepalli2021harmonically,santra2022gap,kethepalli2022edge}.

In this section, we describe a macroscopic procedure and construct a field theory adapted appropriately for our models. We start with the partition function
\begin{align}
    Z_{\beta}(N) &= \int_{-\infty}^{\infty} \prod_{i = 1}^N dp_i \int_{-\infty}^{\infty} {\bf dx}_N {\rm exp}(-\beta H(\{x_i, p_i\})),~\text{where}\notag\\
    \int_{w}^{z}~{\bf dx}_N &\equiv \int_{w}^{z}dx_1\int_{x_1}^{z}dx_2\ldots\int_{x_{N-1}}^{z}dx_N\label{def:dx}.
\end{align}
Since the position and momentum variables are uncoupled, the partition function reduces to
\begin{align}
    Z_{\beta}(N) = Z^{(\rm \mathcal{K})}_{\beta}(N)~Z^{(\rm \mathcal{C})}_{\beta}(N),    
\end{align}
where the configurational contribution to the partition function is given by
\begin{align}\label{Eq:partx}
    Z^{(\rm \mathcal{C})}_{\beta}(N) = &\int_{-\infty}^{\infty} \notag {\bf dx}_N ~ \times \\&{\rm exp}\Big(-\beta\Big[\sum_{i=1}^N U_{\delta}(x_i) + \frac{1}{2}\sum_{i=1}^N\sum_{\substack{j=1 \\ j\neq i}}^{N}V(x_i-x_j) \Big]\Big),
\end{align}
and contribution due to kinetic terms is
\begin{align}
    Z^{(\rm \mathcal{K})}_{\beta}(N) &= \left(\dfrac{2\pi}{\beta}\right)^{N/2}.
\end{align}
Performing the multiple integrals in Eq. \eqref{Eq:partx} is a hard problem. 
However, for (i) short-range repulsive interactions that diverge at vanishing separation, and (ii) slowly varying confining potentials, one can approximate the full partition function as follows.
\begin{figure*}[htb]
	\centering
	{\includegraphics[width=17.85cm]{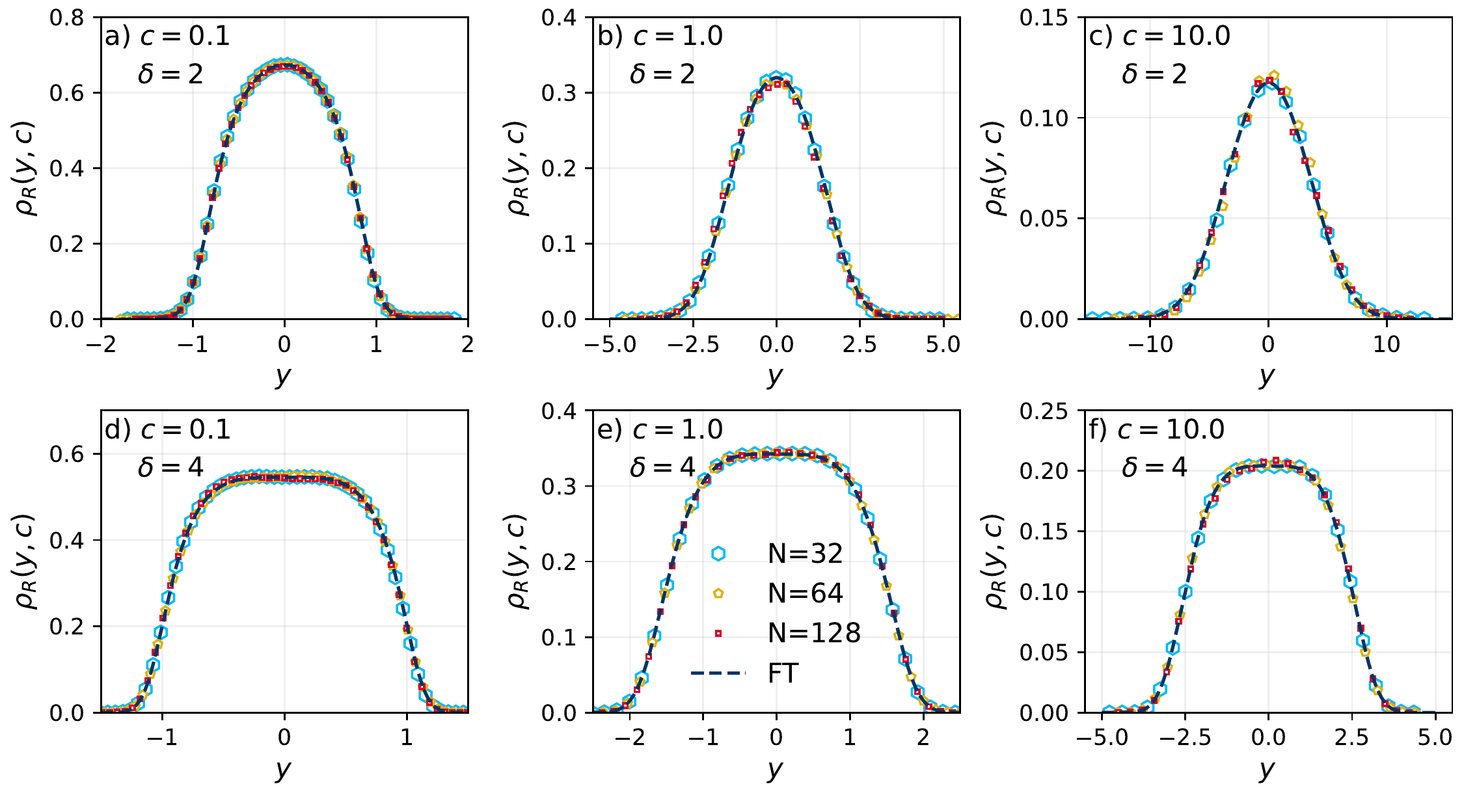}}
	\caption{Comparison of scaled equilibrium density profiles $\rho_R(y,c)$, obtained from Monte-Carlo simulations with field theory [Eq.~\eqref{hrd:s-chem}] denoted by `FT', for the HR model with [(a)-(c)]  quadratic trap ($\delta = 2$)  and [(d)-(f)] quartic trap ($\delta = 4$). We show Monte-Carlo data for three values of $c$: $c= 0.1$, $c = 1.0$ and $c = 10.0$, for $N = 32, 64, 128$. Here the scaled variables are related to the unscaled variables as $y=x/N$ and $ c = T/N^{\delta}$ as given in Eq.~\eqref{eqn:rod-scal}. }
	\label{fig:HRods}
\end{figure*}
We divide the system into $N_b$ subsystems, as shown in Fig.~\ref{fig:schematic}, where each subsystem $s$ contains a large number of particles $n_s$. Note that the size of each subsystem, denoted by $\Delta$, is small enough compared to the actual size of the gas and large enough to contain many particles such that the change in potential energy between two successive boxes is smaller than thermal energy $T$ i.e., $|V_{\rm ext}(x_{s+1})-V_{\rm ext}(x_{s})|<T$.
The particles in each subsystem experience an effective constant potential that depends on the location of the subsystem $x_s$ inside the trap. The partition function $Z^{(\rm \mathcal{C})}_{\beta}(N)$ in Eq.~\eqref{Eq:partx} can be approximated (in the thermodynamic limit) as the product of the partition functions of these boxes,
\begin{align}\label{eqn:zprod}
    Z^{(\rm \mathcal{C})}_{\beta}(N) &\approx {\rm exp}\left(\sum_{s =1}^{N_b} \log \big[ Z_{\beta}(n_s, x_s, \Delta)\big]\right),
\end{align}
where the partition function of the $s^{\rm th}$ subsystem of size $\Delta$, centered around $x_s$ containing $n_s$ particles is given by
\begin{align}
Z_{\beta}(n_s, x_s, \Delta) =\int_{x_s-\frac{\Delta}{2}}^{x_s+\frac{\Delta}{2}}& {\bf dx}_{n_s} \prod_{i=1}^{n_s} \exp\Big(-\beta U_{\delta}(x_i)\Big)\notag \\&\prod_{\substack{i, j=1 \\ j\neq i}}^{n_s} {\rm exp}\Bigg(-\beta \left[\frac{1}{2} V(x_i-x_j)\right]\Bigg)
\end{align}
The free energy per particle in the $s^{\rm th}$ box is given by
\begin{align}\label{def:fint}
 f\left(x_s, \beta\right) = -\frac{1}{\beta n_s}\log \big[Z_{\beta}(n_s, x_s, \Delta)\big].
\end{align}
We convert the summation in Eq.~\eqref{eqn:zprod} over subsystem index $s$ to an integral over $x$ and get [see Appendix~\ref{AppI}]
\begin{align}\label{eqn:mathcalF}
	\mathcal{F}[\rho(x), \beta] = \int_{-\infty}^{\infty} dx~~ \rho(x) f \left(x, \beta\right).
\end{align}
where $\rho(x)$ is the density of particles at position $x$. The free energy per particle $f(x_s, \beta)$ defined in Eq.~\eqref{def:fint} can be computed from the partition function of the subsystem. As mentioned earlier, we assume that the subsystem size $\Delta$ is small enough such that all the $n_s$ particles with position $x_i$ (where $i = 1,2..,n_s$), experience a constant potential $U_{\delta}(x_i) \approx U_{\delta}(x_s)$. The subsystem partition function can then be approximated as
\begin{align}\label{part:subsystem}
Z_{\beta}(n_s, x_s, \Delta) &\approx \notag {\rm exp} \Big(-\beta n_s U_{\delta}(x_s)\Big)\times\\ & \Bigg[\int_{x_s-\frac{\Delta}{2}}^{x_s+\frac{\Delta}{2}} {\bf dx}_{n_s} \prod_{\substack{i, j=1 \\ j\neq i}}^{n_s} {\rm exp}\Big(-\beta \left[\frac{1}{2} V(x_i-x_j)\right]\Big)\Bigg].
\end{align}
Note that, in Eq.~\eqref{part:subsystem}, the $x_{i}$ is a running integration variable not to be confused with the position of the center of the subsystem $x_s$. The contribution to the free energy per particle from the $s^{\rm th}$ box is written as
\begin{align}
	f(x_s, \beta) = U_{\delta}(x_s) + f_{\rm int}(x_s, \beta),
\end{align}
where
\begin{align}\label{free:int-sub}
f_{\rm int}(x_s, \beta) & \notag = -\frac{1}{\beta n_s} \times \\&\log\left(\int_{0}^{\Delta} {\bf dx}_{n_s} \prod_{\substack{i, j=1 \\ j\neq i}}^{n_s}{\rm exp}\Big[-\frac{\beta}{2} V(x_i-x_j)\Big]\right).
\end{align}
From Eq.~\eqref{free:int-sub} one can further rewrite $f_{\rm int}(x_s, \beta) \equiv f_{\rm int}\big(\rho(x_s), \beta\big)$.
Furthermore, using Eq.~\eqref{free:int-sub} we can rewrite Eq.~\eqref{eqn:mathcalF} as [see Appendix~\ref{AppI}]
\begin{align}\label{free:sys}
\mathcal{F}[\rho, \beta] = \int_{-\infty}^{\infty} dx~\rho(x) \Big\{U_{\delta}(x) + f_{\rm int}\big(\rho(x), \beta\big)\Big\}.
\end{align}
For the HR and the HC models the explicit forms of the free energy are derived in Appendix~\ref{appendix:free_rod} and Appendix~\ref{appendix:free_calo} respectively.
\begin{figure*}[htb]
	{\includegraphics[scale=0.7]{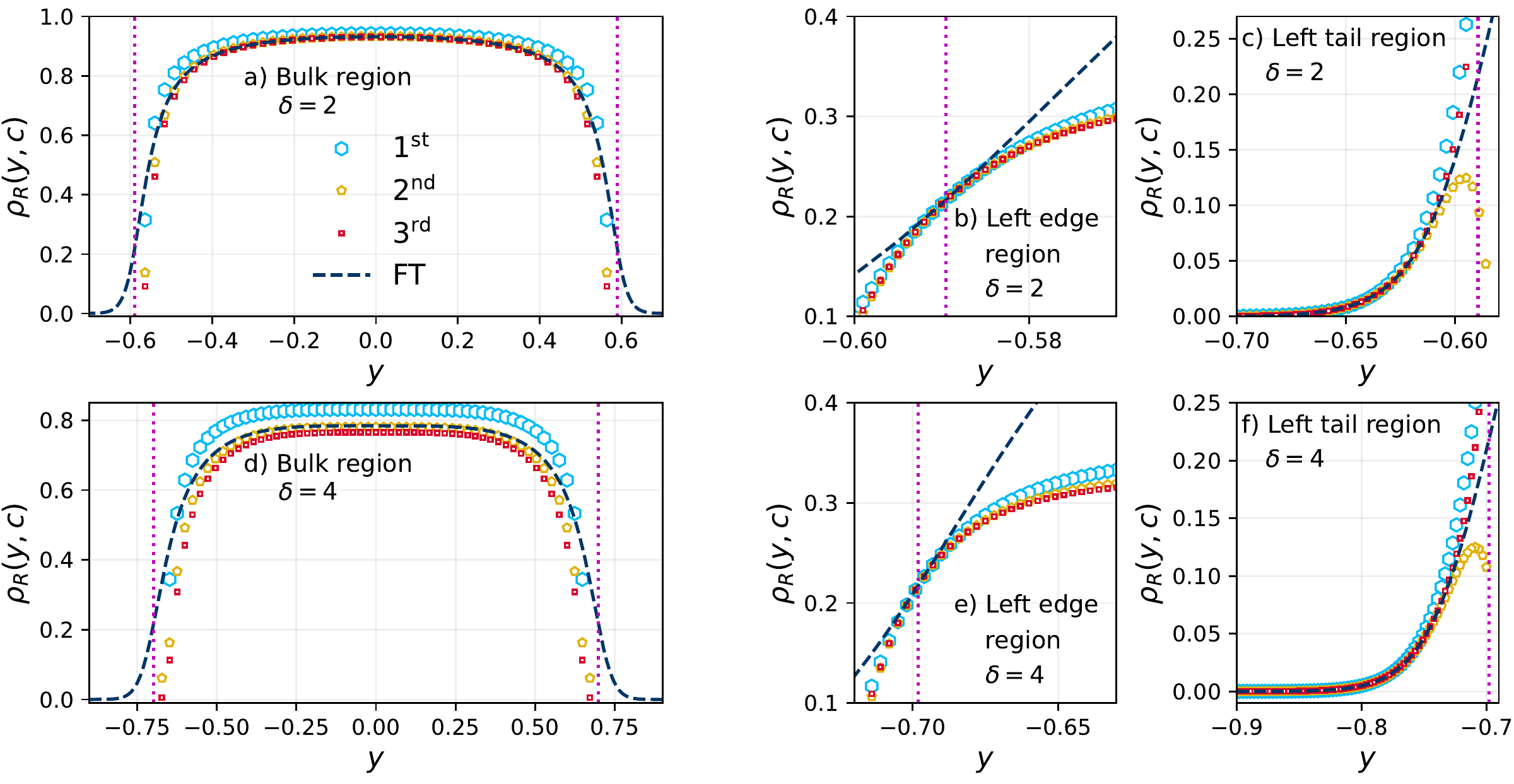}}
	\caption{A comparison of the asymptotic densities up to third iteration (see Appendix.~\ref{appendix:hrd_low_c}) with the densities obtained from the numerical solution of Eq.~\eqref{hrd:s-chem}, denoted by `FT', at low temperature $c = 0.01$ for the HR model in [(a)-(c)] quadratic trap ($\delta=2$) and [(d)-(f)] quartic trap ($\delta = 4$).
	show the densities for hard rods confined to quadratic trap ($\delta = 2$) . Here [(a),(d)] show the bulk ($|y|<y_c-O(c)$), [(b),(e)] edge ($|y-y_c| \lesssim O(c)$) and [(c),(f)] tail ($|y|>y_c+O(c)$) regions. The vertical dotted line represent the position $y = y_c$ given in Eq.~\eqref{hrd:yc_main} and this determines these three regions.}
	\label{fig:hrod_dens_low}
\end{figure*}
The average thermal density can then be computed by extremizing the free energy in Eq.~\eqref{free:sys} with the constraint that the density is normalized
\begin{align}\label{norm}
	\int_{-\infty}^{\infty}dx~\rho(x) = N.
\end{align}
In the next section, we compute these densities for both HR and HC models and compare them with MC simulations. 



\section{Results from field theory and comparison with Monte-Carlo simulations}
\label{density} 
In this section, we adapt the field theory formalism discussed in Sec.~\ref{ftf} for the case of hard rods (HR) and hyperbolic Calogero (HC) model to compute the free energy. We extremize the obtained free energy along with the constraint that the density is normalized and this yields the average density profile. 
\subsection{Hard rods model}\label{hrd}
For the HR model, the contribution due to interaction to the free energy per particle at position $x$ is given by (see Appendix~\ref{appendix:free_rod})
\begin{align}\label{hrd:fint}
f_{\rm int}(\rho(x), \beta) = -\frac{1}{\beta} \log\left(\frac{1-  a~ \rho(x)}{\rho(x)}\right)+\frac{1}{\beta}.
\end{align}
Using Eq.~\eqref{hrd:fint} in Eq.~\eqref{free:sys} we get the free energy for the HR model [ignoring the density independent term $1/\beta$ in Eq.~\eqref{hrd:fint}]
\begin{align}\label{free:hrd-sys1}
    \mathcal{F}_R \left[\rho(x), \beta\right] = \int_{-\infty}^{\infty}~dx~\rho(x) \Bigg[ U_{\delta}(x) -\frac{1}{\beta} \log\left(\frac{1-  a~\rho(x)}{\rho(x)}\right) \Bigg].
\end{align}
The free energy in Eq.~\eqref{free:hrd-sys1} is super-extensive i.e., $\mathcal{F}_R[\rho(x), \beta]\sim O(N^{\delta+1})$ since the ground state energy of $N$ hard rods in a confining potential $U_{\delta}(x)\sim x^{\delta}$ scales as $N^{\delta+1}$. Therefore, the average thermal density $\rho^*(x, T)$ can be computed via saddle point approximation~\cite{agarwal2019harmonically}. This amounts to extremizing the free energy along with the normalization constraint
\begin{align}\label{hrd:saddle}
	\frac{\delta}{\delta \rho(x)}\mathcal{F}_R[\rho(x), \beta]\Bigg|_{\rho(x) = \rho^*(x, T)} = \mu_N(\beta),
\end{align}
where the chemical potential $\mu_N(\beta)$ is temperature dependent and can be extracted from normalization condition given in Eq.~\eqref{norm}. Using Eq.~\eqref{free:hrd-sys1} in Eq.~\eqref{hrd:saddle} we get 
\begin{align}\label{eqn:chem_rod}
    \mu_{N}(\beta) =& U_{\delta}(x)\notag\\& - T \Bigg[\log\left(\frac{1-a~\rho^*(x, T)}{\rho^*(x, T)}\right)  - \frac{1}{1-a~ \rho^*(x, T)}\Bigg].
\end{align}
To obtain the system size dependence of the density profile, we define 
\begin{align}\label{hrd:rhoN}
\rho_N(x, T) = \frac{1}{N}\rho^*(x, T),
\end{align}
such that 
\begin{align}
\int_{-\infty}^{\infty} ~dx~\rho_N(x, T) = 1.
\end{align}
Using Eq.~\eqref{hrd:rhoN}, Eq.~\eqref{eqn:chem_rod} can then be expressed as 
\begin{align}\label{hrd:chem}
    \mu_{N}(\beta) =& U_{\delta}(x) \notag \\ &- T \Bigg[\log\left(\frac{1-a~N \rho_N(x, T)}{N \rho_N(x, T)}\right) - \frac{1}{1-a~N\rho_N(x, T)}\Bigg].
\end{align}
To extract the system size ($N$) and temperature ($T$) dependence of the density $\rho_N(x, T)$ we substitute the following scaling form ansatz
\begin{align}\label{dens:hrd-scal}
    \rho_N(x, T) &= N^{-\alpha_R}~\rho_R \left(y,c\right), \quad ~\mu_N(\beta) =  N^{\lambda_R}~\mu_R(c),
\end{align}
with the scaled variables given by
\begin{align}
    y& = \frac{x}{N^{\alpha_R}}, \quad c= \frac{T}{N^{\gamma_R}},  \label{eqn:rod-scal}
\end{align}
in Eq.~\eqref{hrd:chem}. Here $\alpha_R$ and $\gamma_R$ are scaling exponents which are determined by requiring that Eq.~\eqref{hrd:chem} is $N$ independent in the scaled variables. Doing so we get 
\begin{align}
\alpha_R =1,\quad \gamma_R = \delta~~\text{and}\quad \lambda_R = \delta.
\end{align}
The value $\alpha_R = 1$ can be understood from the $O(N)$ extent of the density profile at zero temperature. This leads to $O(N^{\delta+1})$ energy of the system in the ground state. In order for the entropy term to contribute to the free energy one needs to scale the temperature by $N^{\delta}$ implying $\gamma_R = \delta$. Eq.~\eqref{hrd:chem} finally becomes
\begin{align}\label{hrd:s-chem}
    \mu_R(c) = \frac{y^{\delta}}{\delta} -c \Bigg[\log\left(\frac{1-a~\rho_R (y,c)}{\rho_R (y,c)}\right) - \frac{1}{1-a~\rho_R (y,c)}\Bigg].
\end{align}
It is worth noting that the thermal equilibrium properties of  hard rods in an external potential were studied in Ref.~\onlinecite{percus1976equilibrium}. Eq.~\eqref{hrd:s-chem} can be obtained from Eq.~[13] of Ref.~\onlinecite{percus1976equilibrium}, when the density is assumed to vary slowly on the rod length scale $a$. Since Eq. \eqref{hrd:s-chem} is a transcendental equation, it is difficult to obtain an exact solution. We solve Eq.~\eqref{hrd:s-chem} numerically by fixing $\mu_R(c)$ such that the normalization constraint,
\begin{align}\label{hrd:norm}
\int_{-\infty}^{\infty} dy~\rho_R(y,c) =1,
\end{align}
is satisfied. In Fig. \ref{fig:HRods} we show the comparison between the scaled density profile [obtained by solving Eq.~\eqref{hrd:s-chem}] and data from MC simulations (using the standard Metropolis algorithm) for three rescaled temperatures $c = 0.1, 1.0, 10.0$ and three system sizes $N= 32, 64, 128$. We find quite remarkable scaling collapse of the MC data with system size which also agrees with the field theory results for both the quadratic ($\delta = 2$) and quartic ($\delta = 4$) traps.

\begin{figure}
	{\includegraphics[scale=0.75]{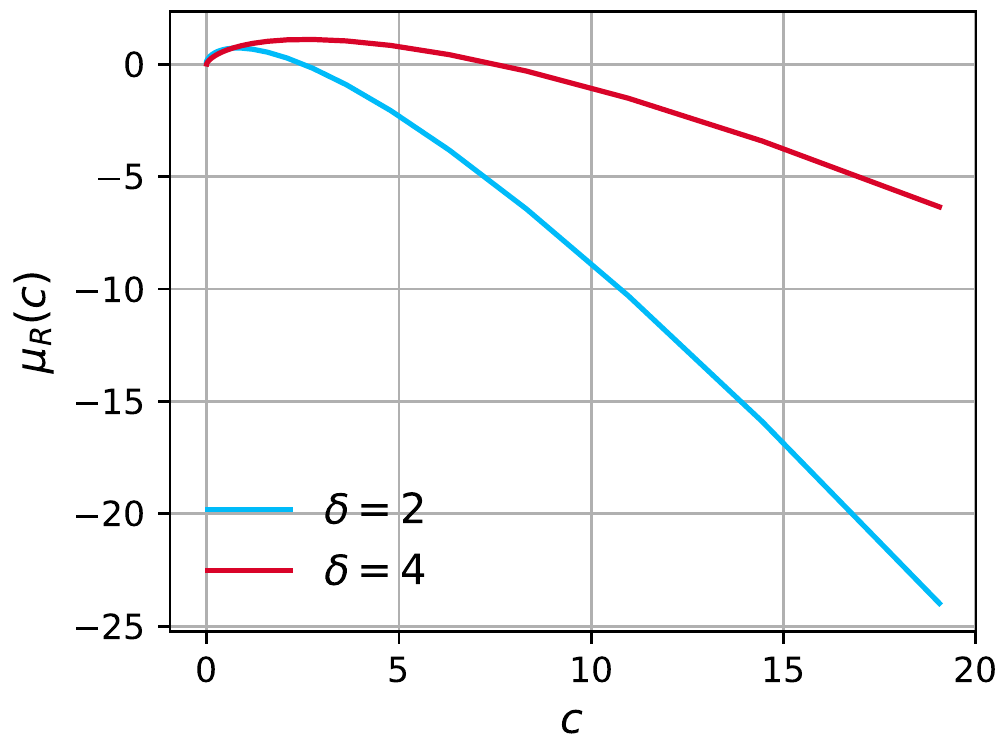}}
	\caption{Chemical potential, $\mu_R(c)$, for the HR model obtained using Eq.~\eqref{hrd:s-chem} and Eq.~\eqref{hrd:norm}, plotted as a function of the rescaled temperature $c$ for quadratic trap with $\delta=2$ (blue) and quartic trap with $\delta = 4$ (red). At large values of $c$ the chemical potential is negative and diverges.}
	\label{fig:rod_mu}
\end{figure}

\begin{figure}
	{\includegraphics[width=7.5cm]{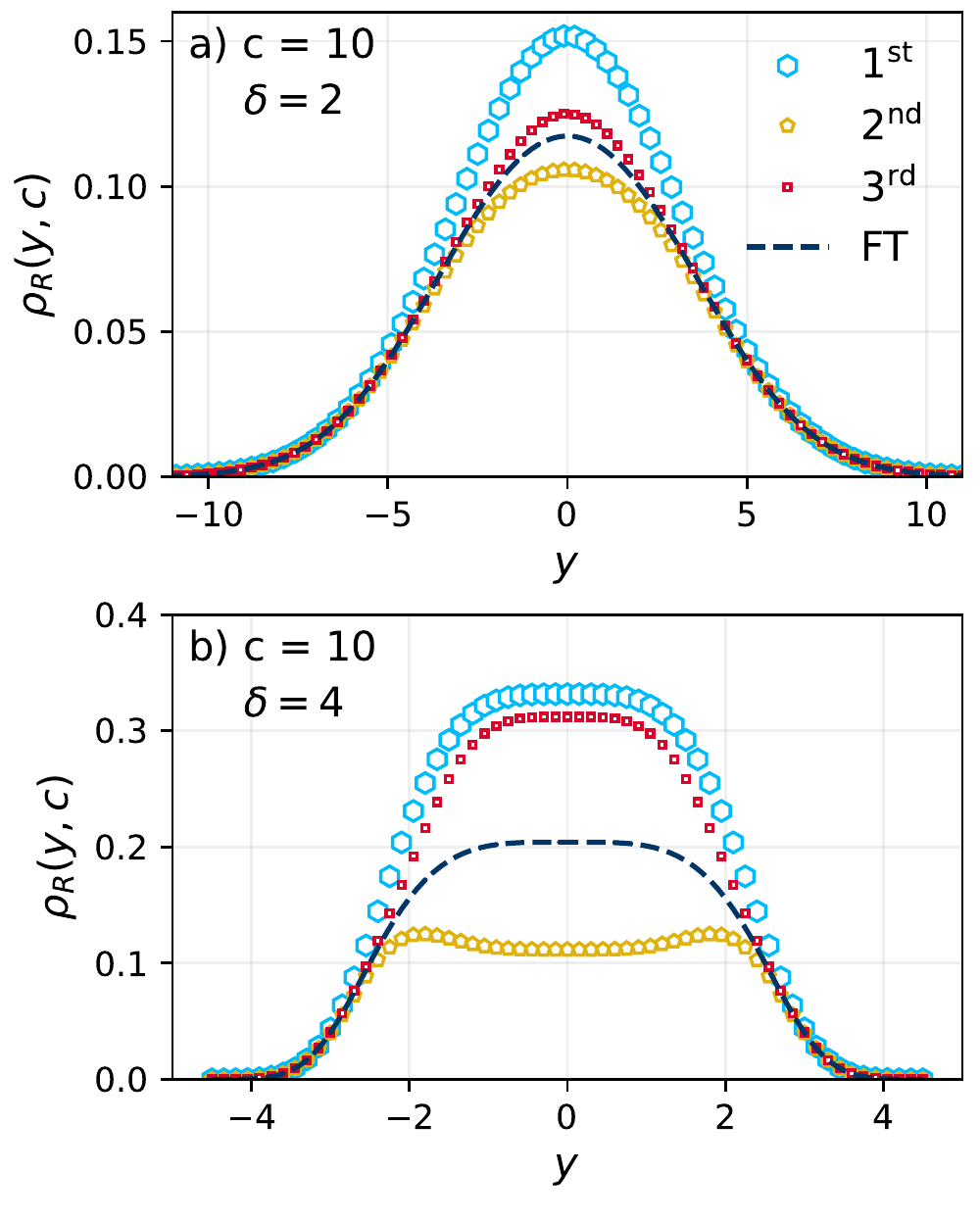}}
	\caption{Comparison of the asymptotic densities up to third order (see Appendix.~\ref{appendix:hrd_high_c}) with the numerical solution of Eq.~\eqref{hrd:s-chem}, denoted by `FT', at high temperature $c = 10.0$ for the HR model confined to (a) quadratic trap ($\delta = 2$) and (b) quartic trap ($\delta = 4$).}
	\label{fig:hrod_dens_high}
\end{figure}

\begin{figure*}[htb]
	\centering
	{\includegraphics[width=18cm]{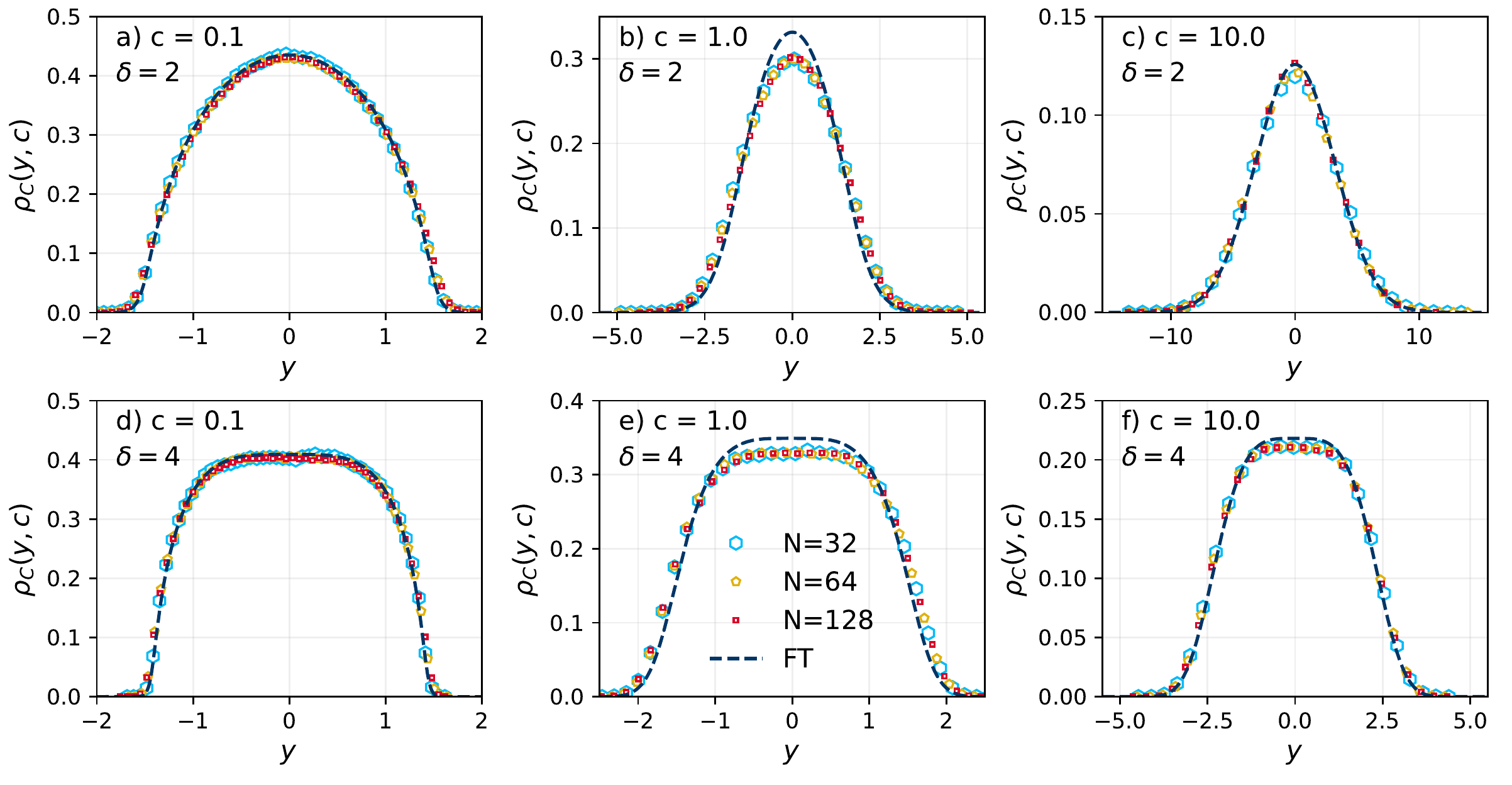}}
	\caption{Comparison of scaled equilibrium density profiles $\rho_C(y,c)$, obtained from Monte-Carlo simulations with field theory [Eq.~\eqref{hcm:scaling-chem2}], denoted by `FT', for the HC model with [(a)-(c)] quadratic trap ($\delta = 2$)  and [(d)-(f)] quartic trap ($\delta = 4$) . We show MC data for three values of $c$: $c= 0.1$, $c = 1.0$ and $c = 10.0$, for $N = 32, 64, 128$. Here the scaled variables are related to the unscaled variables as $y = x/N^{\alpha_C}$ and $c=T/N^{\gamma_C}$, where $\alpha_C$ and $\gamma_C$ are given in Eq.~\eqref{hcm:agl}. }
	\label{fig:HCM}
\end{figure*}

\begin{figure}[t]
	{\includegraphics[scale=0.75]{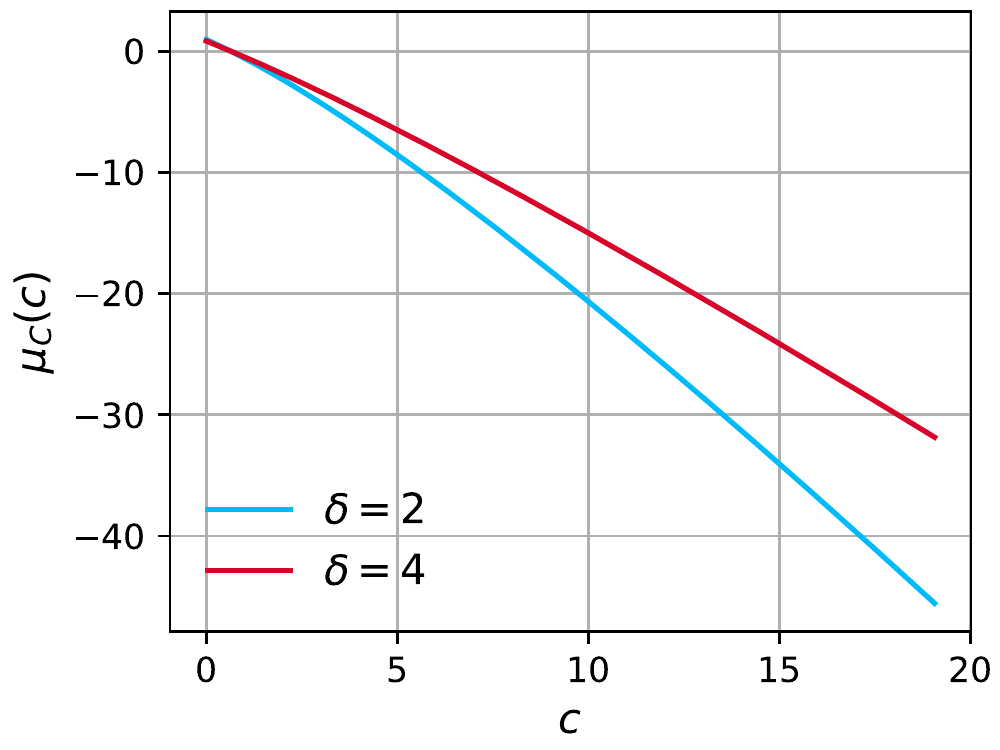}}
	\caption{Chemical potential $\mu_C(c)$ for HC model, computed by using Eq.~\eqref{hcm:scaling-chem2} along with the normalization condition Eq.~\eqref{hcm:normalization}, plotted as a function of the rescaled temperature $c$ for quadratic trap with $\delta=2$ (blue) and quartic trap with $\delta = 4$ (red). At large values of $c$ the chemical potential is negative and diverges.}
	\label{fig:hcm_mu}
\end{figure}

Although explicit analytical solution of the saddle point given in Eq.~\eqref{hrd:s-chem} is highly nontrivial to obtain, one can study the behavior of the density for low $c \ll 1$ and high $c \gg 1$ analytically using asymptotic analysis (see Appendix~\ref{appendix:hrd_low_c}). At zero temperature the hard rods have a density profile given by

\begin{align}\label{hrd:zerotemp_main}
	\rho_R(y, 0) = 
	\begin{cases}
		\frac{1}{a}&~\text{for}~ y\leq \left|\frac{a}{2}\right|\\
		0 &~\text{for}~ y>\left|\frac{a}{2}\right|.
	\end{cases}
\end{align}
The density profile at low temperatures can then be approximated as
\begin{align}\label{hrd:lowc}
	\rho_R(y,c) \stackrel{ c\ll 1}{\approx} \rho_R(y,0)+\rho_1(y,c),
\end{align}
where the deviation from the zero temperature density up to first iteration (see Appendix~\ref{appendix:hrd_low_c} for more details) is given by
\begin{align}\label{hrd:dens-corr-low}
	\rho_1(y,c) \approx 
	\begin{cases}
		&-\frac{1}{a}\frac{c~\delta}{\left(y_c^{\delta}-y^{\delta}\right)}~\text{for}~|y|<y_c-O(c)\\
		&\rho_R^*(1-a~\rho_R^*)^2\times \\&\left(1- \exp\left[-\frac{y_c^{\delta} -y^{\delta}}{ c~\delta}\right]\right)~\text{for}~|y-y_c|<O(c)\\
		&\frac{1}{e}\exp\left(\frac{y_c^{\delta} -y^{\delta}}{ c~\delta}\right)~\text{for}~|y|>y_c+O(c).
	\end{cases}
\end{align}

\noindent
Here $y_c$ is the position at which the term in the parenthesis of Eq.~\eqref{hrd:s-chem} changes sign and is given by
\begin{align}\label{hrd:yc_main}
	y_c = ( \mu_R\delta)^{\frac{1}{\delta}}.
\end{align}
The density at $y = y_c$ is denoted by 
\begin{align}
	\rho^*_R = \rho_R(y_c,c).
\end{align}
Note that $y_c$ given in Eq.~\eqref{hrd:yc_main} determines the three regions (see Appendix ~\ref{appendix:hrd_low_c})
	(i) bulk region where $|y|<y_c-O(c)$,
	(ii) edge region where $|y-y_c|\lesssim O(c)$,	
and	(iii) tail region where $|y|>y_c+ O(c)$,	
which are displayed in  Eq.~\eqref{hrd:dens-corr-low}. The higher order corrections have also been computed and are presented in Appendix~\ref{appendix:hrd_low_c}.  The expression Eq.~\eqref{hrd:lowc} is verified with numerical solution of Eq.~\eqref{hrd:s-chem} for $c=0.01$ in Fig.~\ref{fig:hrod_dens_low} showing the three regions. For this comparison the value of the chemical potential $\mu_R(c)$ is taken from Fig.~\ref{fig:rod_mu}. Note that in Fig.~\ref{fig:rod_mu} the behavior of $\mu_R(c)$ is non-monotonic: it increases initially as $c$ is increased from zero and thereafter decreases. This nonmonotonicity can be explained by noting that at smaller $c$, particles can only be added to the edges of the system which requires more energy (owing to the confining potential), without a large increase in entropy.  Hence $\mu_R(c)$ increases initially. However, for larger $c$, the gas expands and this opens up gaps, larger than the size of the rods, between the particles in the bulk of the system. Consequently, one can easily add an extra rod with a small energy cost and a large entropy gain, essentially lowering the free energy change. Hence $\mu_R(c)$ decreases with $c$ for larger $c$.

In the high temperature regime ($c \gg 1$) the spread of the gas increases which in turn dilutes the gas i.e., $\rho_R(y, c) \ll 1$. Using this low density approximation in Eq.~\eqref{hrd:s-chem}, we obtain the approximate analytical expression of the density profile (up to the first iteration), given by  (see Appendix~\ref{appendix:hrd_high_c})
\begin{align}\label{hrd:highc}
	\rho_R(y,c) &\stackrel{ c\gg 1}{\approx} \frac{1}{e}\exp\left(\frac{\mu_R(c)}{c}-\frac{y^{\delta}}{c \delta }\right),
\end{align}
where the chemical potential $\mu_R(c)$ is obtained numerically by solving Eq.~\eqref{hrd:s-chem} along with the normalization condition [Eq.~\eqref{hrd:norm}] as shown in Fig.~\ref{fig:rod_mu} and $e=2.71828$ is the Euler's number. We can obtain higher order terms of the density by also considering subdominant corrections originating due to the presence of interaction as shown in Appendix~\ref{appendix:hrd_high_c}. The expression Eq.~\eqref{hrd:highc} and the subdominant corrections (up to third order) are verified with the numerical solution of Eq.~\eqref{hrd:s-chem} for both traps in Fig.~\ref{fig:hrod_dens_high} for $c = 10.0$.


\subsection{Hyperbolic Calogero model}\label{hcm}
\begin{figure*}
	{\includegraphics[scale=0.65]{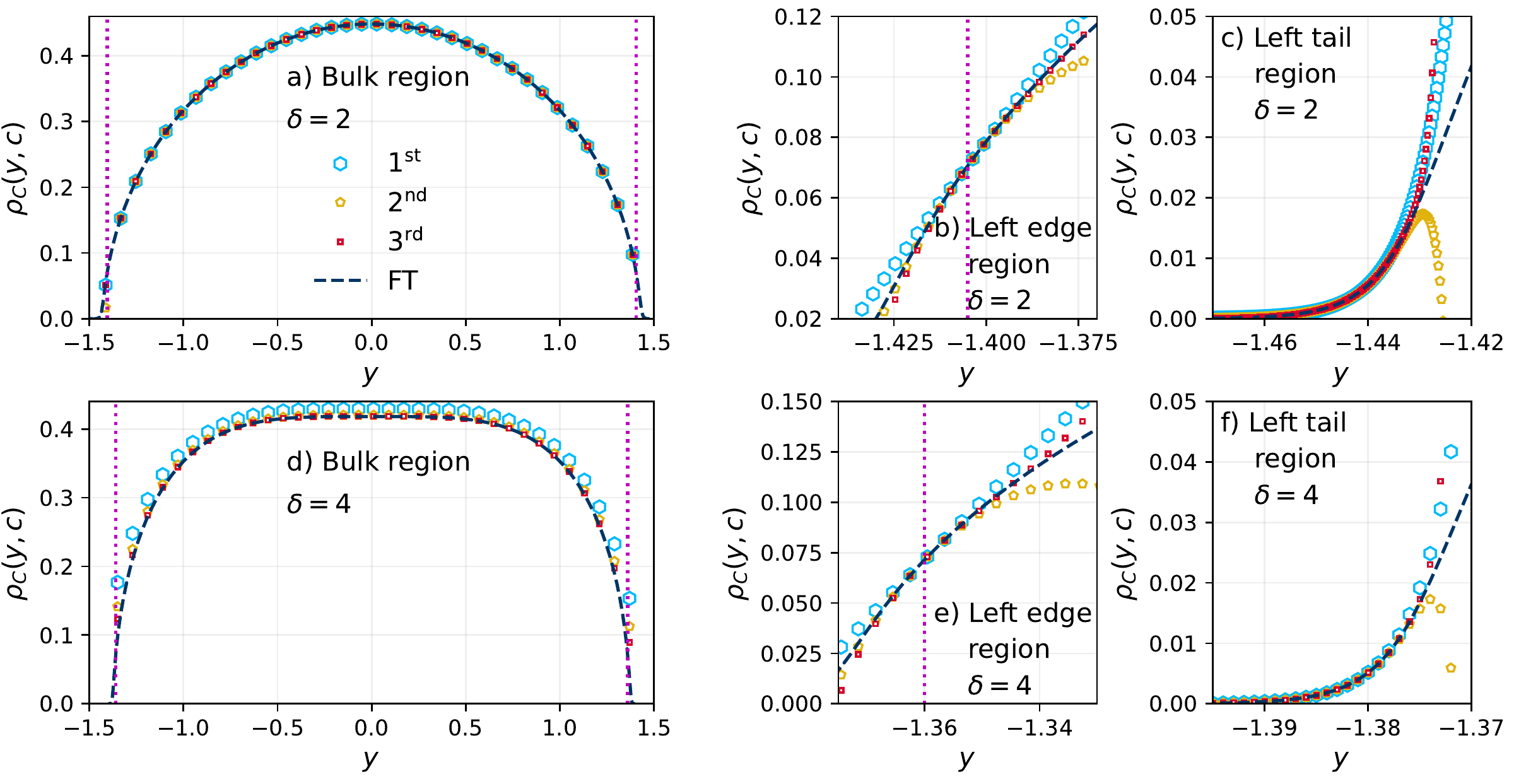}}
	\caption{A comparison of the asymptotic densities up to third order (see Appendix.~\ref{appendix:hcm_low_c}) with the densities obtained from the numerical solution of Eq.~\eqref{hcm:scaling-chem2}, denoted by `FT', at low temperature $c = 0.01$ for the HC model. Here we show the densities for HC model confined to [(a)-(c)] quadratic trap ($\delta = 2$) and [(d)-(f)] quartic trap ($\delta = 4$). Here [(a),(d)] show the bulk ($|y|<y_c$), [(b),(e)] edge ($|y-y_c| \lesssim O(c)$) and [(c),(f)] tail ($|y|>y_c$) regions. The dotted vertical line represent the position $y = y_c$ which determines the three regions. In (c) and (f), as $y = y_c$ falls outside of the domain of the $x$-axis, for the sake of presentation, we do not show the dotted line.}
	\label{fig:hhcm_dens_low}
\end{figure*}
\begin{figure}
	{\includegraphics[scale=0.69]{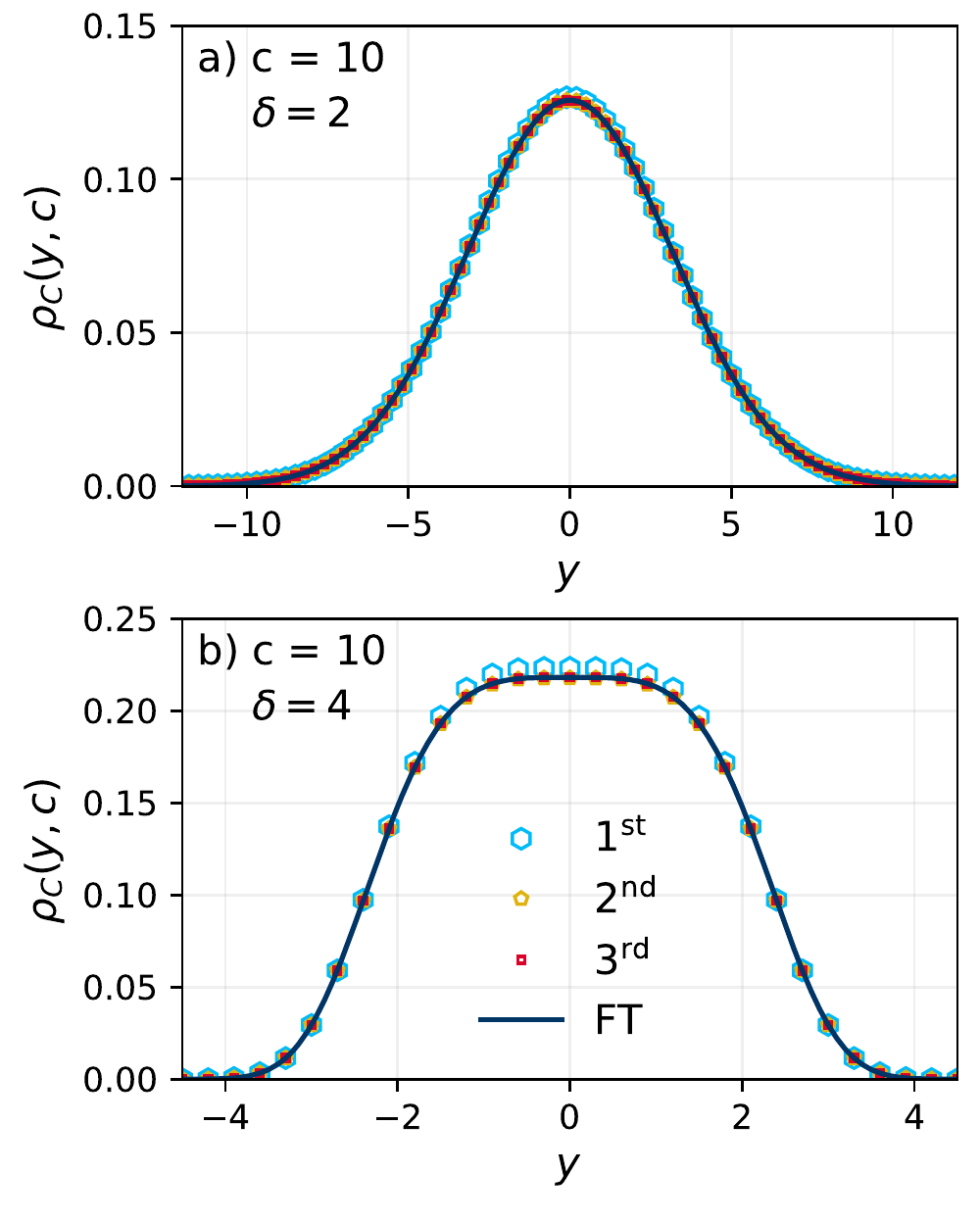}}
	\caption{Comparison of the asymptotic densities up to third order (see Appendix.~\ref{appendix:hcm_high_c}) with the numerical solution of Eq.~\eqref{hcm:scaling-chem2}, denoted by `FT', at high temperature $c = 10.0$ for the HC model confined to (a) quadratic trap ($\delta = 2$) and (b) quartic trap ($\delta = 4$).}
	\label{fig:hhcm_dens_high}
\end{figure}

Unlike the hard rods (HR) model, for the hyperbolic Calogero (HC) model, each particle is coupled to all the other particles. The field theoretic formulation of the hyperbolic Calogero model has been studied ~\cite{gon2019duality}. However, the average thermal density profiles at finite temperatures have not been computed yet. Based on the the approximate scheme outlined in Sec.~\ref{ftf} and the approach described in Refs. \cite{stone2008classical, kumar2020particles, agarwal2019harmonically}, we compute the finite temperature density profiles for the hyperbolic Calogero model below. The free energy in this case is given by  (see Appendix~\ref{appendix:free_calo})
\begin{align}\label{hcm:free_rho}
   & \mathcal{F}_C \left[\rho(x), \beta\right]\notag \\ & = \int_{-\infty}^{\infty}~dx~\rho(x) \left[U_{\delta}(x) + J \zeta(2) \rho(x)^2+ \frac{1}{\beta} \log\big[\rho(x)\big]\right],
\end{align}
where $\zeta(k) = \sum_{n = 1}^{\infty} n^{-k}$ is the Riemann Zeta function. Note that, despite the all-to-all coupling, the contribution to the free energy per particle due to interactions gets renormalized to a local term in the density field, and is given by
\begin{align}
f_{\rm int}(\rho(x), \beta) = J \zeta(2) \rho(x)^2+\frac{1}{\beta}\log \rho(x).
\end{align}
Here $\beta^{-1}\log\rho(x)$ is the contribution due to the configurational entropy. To compute the average thermal density $\rho^*(x, T)$, we extremize the free energy functional given in Eq.~\eqref{hcm:free_rho} along with the normalization condition [Eq.~\eqref{norm}] which gives the chemical potential
\begin{align}\label{hcm:chem}
    \mu_N(\beta) = U_{\delta}(x) + 3 \zeta(2) \rho^*(x, T)^2 + T \Big(1 + \log\big[\rho^*(x, T)\big]\Big).
\end{align}
As in the case of HR model, to obtain a scaling form for the density profile, we use the density normalized to unity $\rho_N(x, T) = \rho^*(x, T)/N$ in Eq.~\eqref{hcm:chem}  and get
\begin{align}\label{hcm:chem-unit-normal}
    \mu_N(\beta) = U_{\delta}(x) &+ 3 \zeta(2) N^2 \rho_N(x, T)^2 \notag \\&+ T \Big(1 + \log \big[N \rho_N(x, T)\big]\Big).
\end{align}
\begin{figure*}[htb]
	\centering
	{\includegraphics[width=18cm]{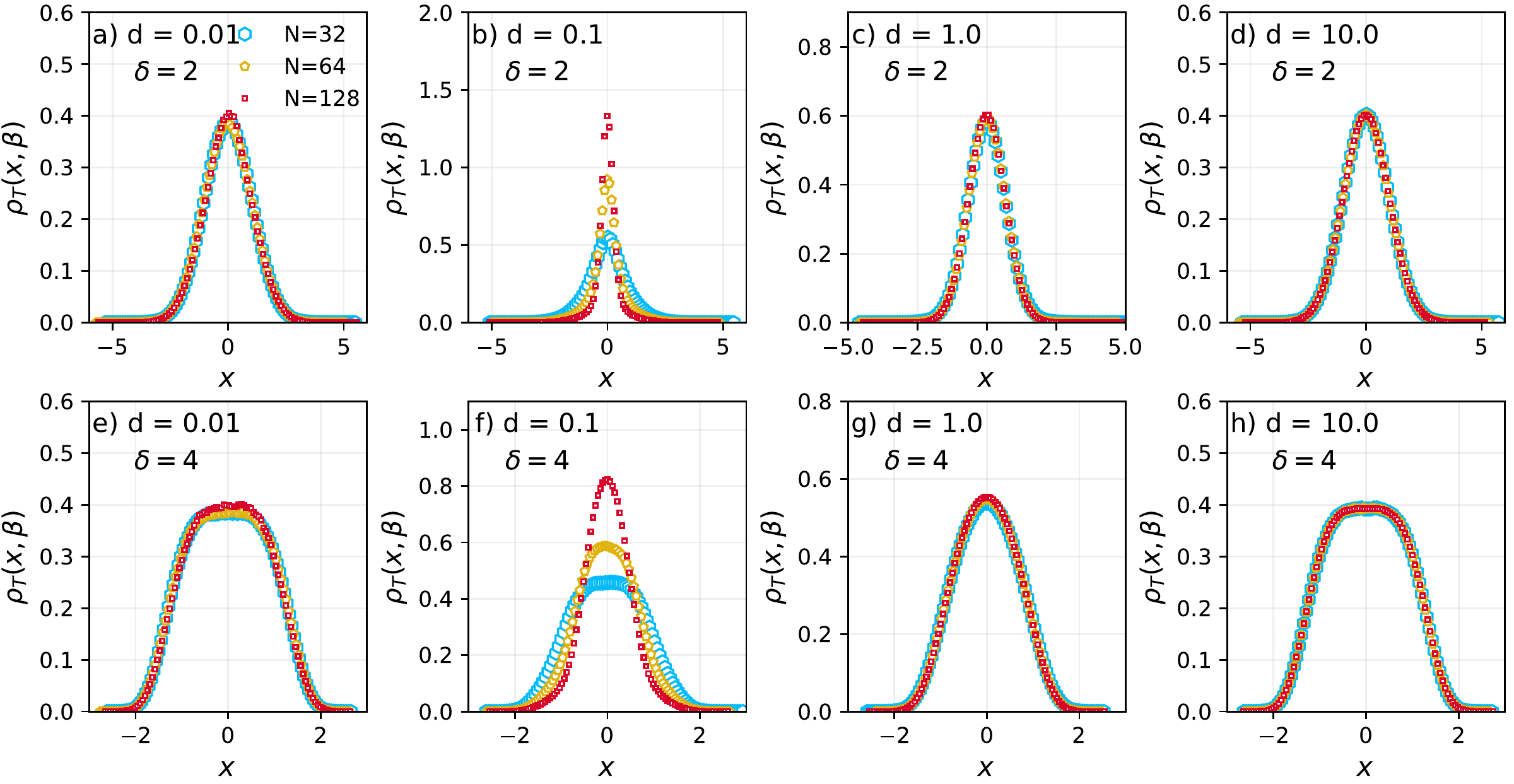}}
	\caption{Equilibrium density profiles obtained from MC simulations of the Toda model when confined to the [(a)-(d)] quadratic trap ($\delta = 2$)  and [(e)-(h)] quartic trap ($\delta=4$) , for different values of the interaction length scale, $d = 0.01, 0.1, 1.0$ and $10.0$. Except for $d = 0.1$ [(b) and (f)], density collapse for different system sizes $N=32, 64, 128$ is observed for all the other values of $d$.}
	\label{fig:Toda}
\end{figure*}
We can extract the system size ($N$) and temperature ($T$) dependence of the density $\rho_N(x, T)$ by substituting the scaling form ansatz
\begin{align}\label{dens:hcm-scal}
    \rho_N(x, T) = N^{-\alpha_C} \rho_C \left(y, c\right), \quad \mu_N(\beta) = \mu_C(c) N^{\lambda_C},
\end{align}
with the scaled variables
\begin{align}
	y = \frac{x}{N^{\alpha_C}}, \quad c = \frac{T}{N^{\gamma_C}},
\end{align}
in Eq.~\eqref{hcm:chem-unit-normal}. Here $\alpha_C$ and $\gamma_C$ are scaling exponents which are determined by requiring that Eq.~\eqref{hcm:chem-unit-normal} is $N$ independent for large-$N$ and depends only on the scaling variables. Doing so, we get 
\begin{align}\label{hcm:agl}
	\alpha_C = \frac{2}{2+\delta}, \quad \gamma_C = \frac{2 \delta}{2+\delta} \quad \text{and} \quad \lambda_C = \frac{2 \delta}{2+\delta}.
\end{align}
The value $\alpha_C = 2/(2+\delta)$ can be understood from the $O(N^{\frac{2}{2+\delta}})$ extent of the gas at zero temperature~\cite{agarwal2019harmonically}. This leads to $O(N^{\alpha_C\delta+1})$ energy of the system in the ground state. In order for the entropy term to contribute to the free energy one needs to scale the temperature by $N^{\alpha_C\delta}$ implying $\gamma_C = \alpha_C\delta$.
Eq.~\eqref{hcm:chem-unit-normal} finally becomes
\begin{align}\label{hcm:scaling-chem2}
    \mu_C(c) = \frac{y^{\delta}}{\delta} + 3 \zeta(2)\rho_C(y,c)^2 + c \log\rho_C(y,c).
\end{align}
We solve Eq.~\eqref{hcm:scaling-chem2} numerically by fixing $\mu_C(c)$ such that the normalization constraint, 
\begin{align}\label{hcm:normalization}
\int_{-\infty}^{\infty} dy~\rho_C(y,c) =1,
\end{align}
is satisfied. In Fig.~\ref{fig:HCM}, we show the comparison between the scaled density profile obtained by solving Eq.~\eqref{hcm:scaling-chem2} and data from MC simulations for $c=0.1, 1.0, 10.0$. We observe good agreement albeit with some small discrepancies, the origin of which is not understood clearly at present. The value of the chemical potential $\mu_C(c)$ is obtained as a function of $c$ by numerically solving Eq.~\eqref{hcm:scaling-chem2} subject to the normalization condition Eq.~\eqref{hcm:normalization}, which is shown in Fig.~\ref{fig:hcm_mu}. Unlike the HR model, we find that $\mu_C(c)$ decreases monotonically in this case.
\begin{figure*}[htb]
	\centering
	{\includegraphics[scale=0.8]{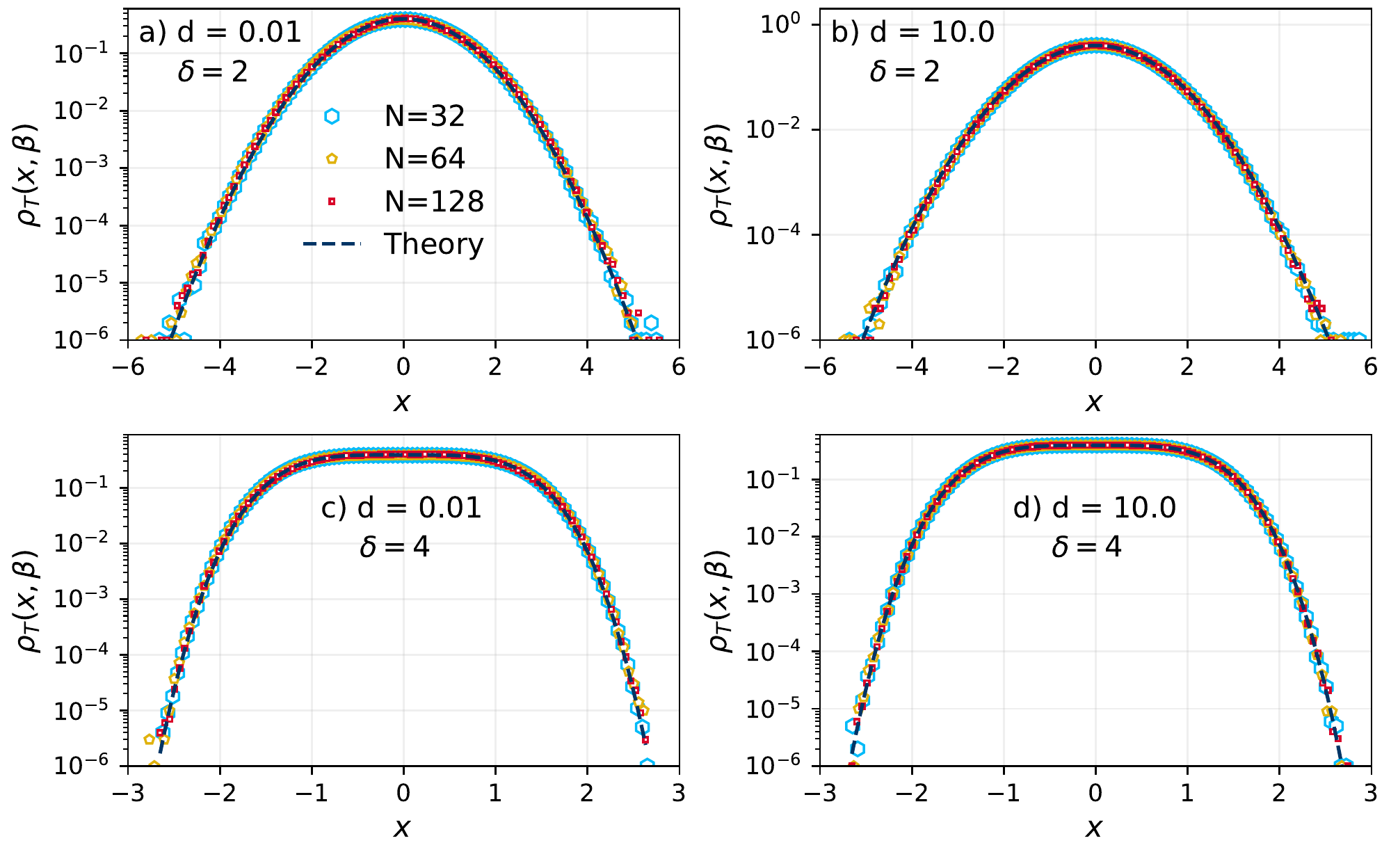}}
	\caption{Equilibrium density profiles obtain from MC simulations for the Toda model confined to [(a),(b)] quadratic trap ($\delta = 2$)  and [(c),(d)] quartic trap ($\delta=4$), for [(a), (c)] small $d = 0.01$  and [(b),(d)] large $d = 10.0$. In both these limits of $d$, the density profile fits well with the form $\rho_N(x) \sim \exp[-\beta \frac{x^{\delta}}{\delta}]$ given in Eq.~\eqref{toda:limit} which is denoted by `Theory'. Here $\beta = 1$ and $N = 32, 64, 128$.}
	\label{fig:Toda_a_limits}
\end{figure*}

\begin{figure}[htb]
	\centering
	{\includegraphics[scale=0.8]{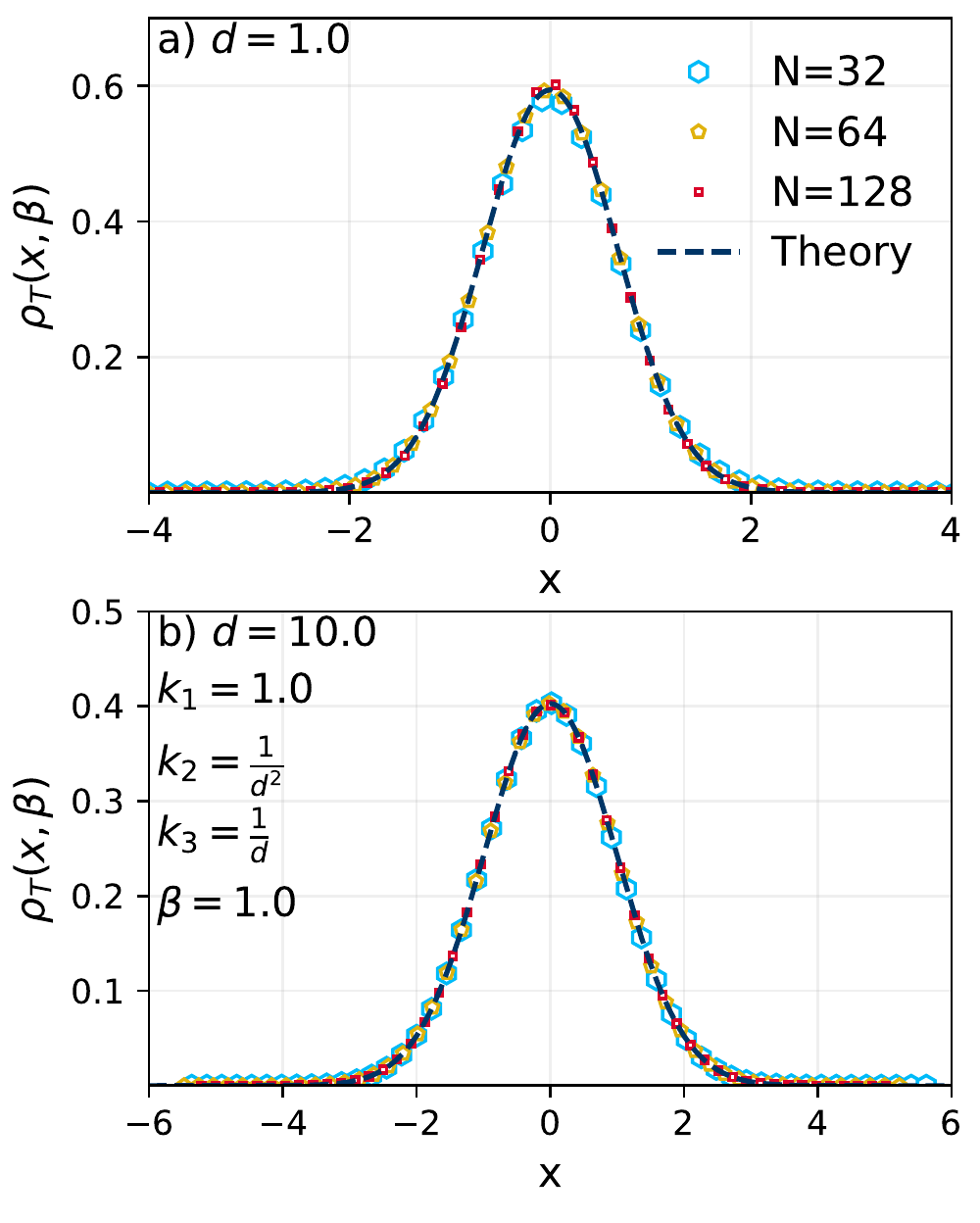}}
	\caption{Comparison of equilibrium density profiles, obtained using MC simulations, with theory Eq.~\eqref{eqn:hrmcdensity} for the nearest neighbor harmonic chain confined to quadratic trap, with the parameters $k_1 = 1.0$, $k_2 = 1/d^2$ and $k_3 = 1/d$ with different values of (a) $d = 1.0$ and (b) $d = 10.0$.}
	\label{fig:harmvstoda}
\end{figure}
Similar to the HR case, obtaining the exact solution of Eq. \eqref{hcm:scaling-chem2} is highly nontrivial for an arbitrary $c$. However, we can study the average thermal density profiles using asymptotic analysis for small $c \ll 1$ and large $c \gg 1$ (see Appendix~\ref{appendix:low_c_C}). At zero temperature, $c = 0$, the density is exactly known and is governed by the interaction term only, since the contribution to the free energy from the entropy is zero. The zero temperature density is given by~\cite{agarwal2019harmonically, kumar2020particles}
\begin{align}\label{hcm0dens}
	\rho_C(y, 0) = 
	\begin{cases}
		A_{\delta}\left(l^{\delta}-y^{\delta}\right)^{\frac{1}{2}} & \text{for} \quad |y|<l\\
		0 & \text{for} \quad |y|>l,
	\end{cases}
\end{align}
where 
\begin{align}\label{hcm:Adelta}
A_{\delta} = \left[3 \delta \zeta(2)\right]^{-\frac{1}{2}}
\end{align}
and the edge of the support of the density is given by
\begin{align}\label{hcm:Edge}
	l = \Big(\mu_C(0) \delta\Big)^{\frac{1}{\delta}} = \left(\frac{\delta}{2 A_{\delta} {\rm B}\left(\frac{1}{\delta}, \frac{3}{2}\right)}\right)^{\frac{2}{2+\delta}},
\end{align}
with
\begin{align}
	{\rm B}(x, y) = \int_{0}^1~dr~r^{x-1}(1-r)^{y-1},
\end{align} 
being the Beta function.  $\mu_C(0)$ in Eq.~\eqref{hcm:Edge} is the scaled chemical potential at zero temperature, obtained by imposing the normalization condition [Eq.~\eqref{hcm:normalization}], and is given by
\begin{align}\label{hcm:muc0}
	\mu_C(0) = \frac{l^{\delta}}{\delta} = \left(\frac{\pi }{2}\right)^{\frac{\delta }{\delta +2}} \left(\frac{\delta ^{-1/\delta } \Gamma \left(\frac{3}{2}+\frac{1}{\delta }\right)}{\Gamma \left(1+\frac{1}{\delta }\right)}\right)^{\frac{2 \delta }{\delta +2}},
\end{align}
where 
\begin{align}
	\Gamma[n] = \int_0^{\infty} ~dx~x^{n-1}e^{-x},
\end{align}
is the Gamma function.

For $c \neq 0$, the entropy starts contributing to the density. As the rescaled temperature is increased from zero, i.e., $c \ll 1$, we can obtain the approximate analytical form of the density profile, as shown in the Appendix~\ref{appendix:hcm_low_c}, which is given by
\begin{align}\label{hcm:lowc}
	\rho_C(y,c) \stackrel{ c\ll 1}{\approx} \rho_C(y,0)+\rho_1(y,c).
\end{align}
Here the deviation from zero temperature density (up to first iteration) is given by
\begin{align}\label{hcm:dens-corr-low}
	\rho_1(y,c) \approx 
	\begin{cases}
	\rho_C(y,0) \times  &\\ \frac{\mu_C(c)-\mu_C(0)-c\log\rho_C(y,0)}{c+6\zeta(2)\rho_C(y,0)^2},&~\text{for}~|y|<y_c-O(c),\\
	\frac{\rho_C^*\left(y_c^{\delta}-y^{\delta}\right)}{\delta\left(c + 6\zeta(2) \rho_C^{*2}\right)}&~\text{for}~|y-y_c|<O(c),\\
	\exp\left(\frac{y_c^{\delta} -y^{\delta}}{c\delta}\right) &~\text{for}~|y|>y_c+O(c),
	\end{cases}
\end{align}
and the higher order corrections are provided in Appendix~\ref{appendix:hcm_low_c}. Similar to the HR model, here $y_c = ( \mu_C(c)\delta)^{\frac{1}{\delta}}$ and $\rho_C^* = \rho_C(y_c,c)$ is the value of the density at $y=y_c$. In Fig.~\ref{fig:hhcm_dens_low}, we find a good agreement between the expression Eq.~\eqref{hcm:lowc} and the numerical solution of Eq.~\eqref{hcm:scaling-chem2} for $c = 0.01$. Note that, for this comparison the value of the chemical potential $\mu_C(c)$ is taken from Fig.~\ref{fig:hcm_mu}, where we recall that $\mu_C(c)$ is obtained by solving Eq.~\eqref{hcm:scaling-chem2} along with the normalization condition [Eq.~\eqref{hcm:normalization}].

As temperature increases the particles spread spatially over a wider region.  Therefore, at high temperatures, $c \gg 1$, the gas becomes dilute i.e. $\rho_C(y,c) \ll 1$. Using this low density approximation in Eq.~\eqref{hcm:scaling-chem2} yields (see Appendix~\ref{appendix:hcm_high_c})
\begin{align}\label{hcm:dens-corr-high}
	\rho_C(y,c) \stackrel{ c\gg 1}{\approx} \exp\left( \frac{\mu_C(c)}{c}-\frac{y^{\delta}}{c\delta}\right),
\end{align}
where $\mu_C(c)$ is obtained numerically from Fig.~\ref{fig:hcm_mu}.
The form of the density in Eq.~\eqref{hcm:dens-corr-high} comes from the entropy which provides the dominant contribution to the density for $c \gg 1$. In Fig.~\ref{fig:hhcm_dens_high}, for $c=10$, we find a good agreement between the approximate expression of the density profile given in Eq.~\eqref{hcm:dens-corr-high} (see Appendix~\ref{appendix:hcm_high_c} for higher correction)  and the numerical solution of the Eq.~\eqref{hcm:scaling-chem2}.
\begin{figure}[htb]
	\centering
	{\includegraphics[scale=0.85]{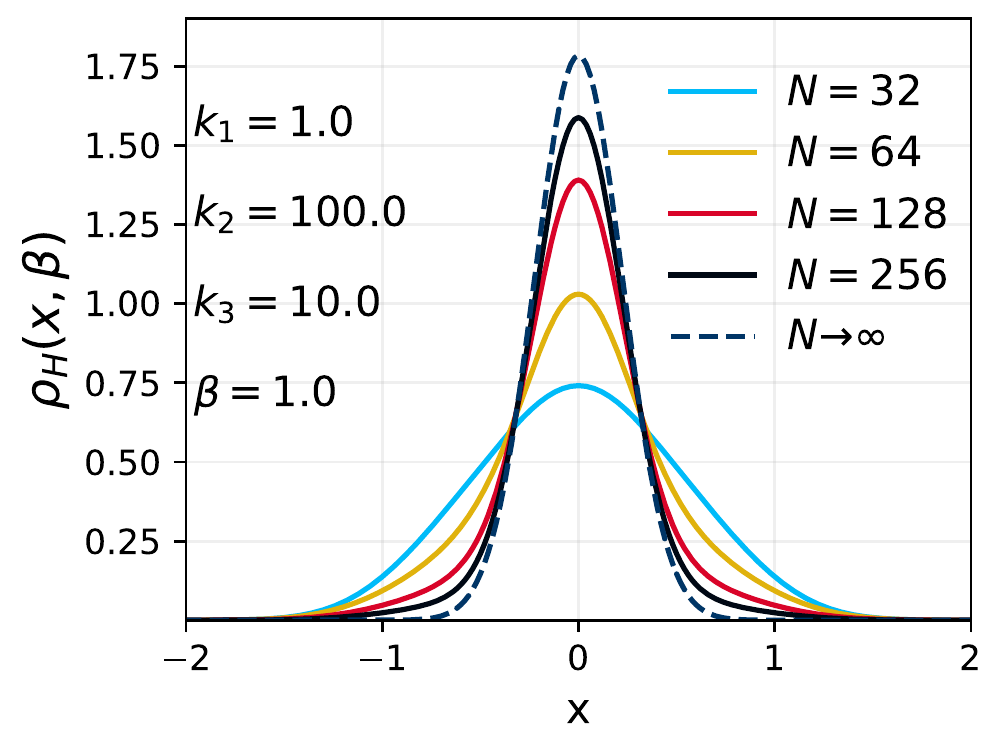}}
	\caption{Analytically computed density profiles [Eq.~\eqref{eqn:hrmcdensity_main}] of the harmonic chain in quadratic trap for $d = 0.1$, $k_1 = 1.0$, $k_2 = 1/d^2$ and $k_3 = 1/d$. For $N=32, 64, 128, 256$ density profile converges very slowly to the Eq.~\eqref{eqn:smalld} of the Appendix~\ref{AppIII}.}
	\label{fig:harm2}
\end{figure}

\subsection{Integrable models with particle-crossing}\label{int}
In the previous sections, we studied hard rods (HR) and hyperbolic Calogero (HC) models which have strong interparticle repulsions that prevent their trajectories from crossing. In this section, we study two models, namely Toda and Harmonic chains, which allow for crossing of particle trajectories due to their weak interparticle repulsion. For these models, the interactions are nearest neighbor. The Toda model takes the form 
\begin{align}\label{eqn:toda}
	V_T(r_i) = J \exp\left(-\frac{r_i}{d}\right), 	
\end{align}
where $r_i = x_{i+1}-x_i$, $J>0$ (we set $J = 1$) is the interaction strength and $d$ determines the length scale of the interaction. The subscript $T$ in the Eq.~\eqref{eqn:toda} stands for the Toda model. For the case of harmonic chain, the interaction takes the form 
\begin{align}\label{eqn:harm}
	V_H(r_i) = \frac{k_2}{2}r_i^2 - k_3r_i,
\end{align}
where $k_2$ and $k_3$ are the interaction strengths. The subscript $H$ in Eq.~\eqref{eqn:harm} stands for the harmonic chain. Note that in the absence of the trap these models are integrable, similar to HR and HC. Furthermore note that the form of the harmonic chain interaction in Eq.~\eqref{eqn:harm} is obtained by expanding the Toda interaction given in Eq.~\eqref{eqn:toda} up to the quadratic order in the nearest neighbor separation $r_i$, with 
\begin{align}
k_2 = \frac{1}{d^2} \quad \text{and} \quad k_3 = \frac{1}{d}.	
\end{align}
This is an analytically tractable model even in the presence of a quadratic trap given by $U_{\delta}(x) = k_1 x^{\delta}/\delta$ with $\delta = 2$. We set $k_1 = 1$.

In contrast to the HR [Sec.~\ref{hrd}] and HC [Se .~\ref{hcm}] models, we find that the field theoretic description fails to describe the equilibrium properties of the trapped Toda and harmonic chain models. The failure of the field theory in this case can be  ascribed to the fact that the particles stay confined to a region of length $\sim O(N^0)$ or smaller, due to the lack of strong repulsion in presence of external confining trap.  To understand this, we consider the behavior of the system at zero temperature. By minimizing the energy one can find the particle positions and it turns out that for both the models a large number [$\sim O(N)$] of particles are confined to a distance of $\sim O(N^{0})$ around the minimum of the external potential. This fact can be understood as follows. Assuming there is a length scale of order $O(N^{\alpha})$, i.e., the particle positions are of order $x_i \sim N^{\alpha}y_i$, we compute the contributions from the potential and interaction energy. The contribution from the external potential is of order $\sum_{i=1}^N U_{\delta}(x_i) \sim O(N^{ \alpha \delta +1} )$. The contribution from the interaction energy for the Toda model scales as $\sum_{i=1}^{N-1} V_{T}(x_{i+1}-x_i) \sim O\big(N \exp \left( {-N^{\alpha}} \right) \big)$ and  for the Harmonic chain it scales as $\sum_{i=1}^{N-1} V_H(x_{i+1}-x_i) \sim O(N^{2\alpha + 1})$. Comparing these energy contributions, we obtain $\alpha = 0$, for both models and for all values of $\delta>0$, implying a length scale of $O(N^0)$. As a consequence the total energy is of $O(N)$. 

As the temperature $T$ is increased, maintaining $T \sim O(1)$, the particles still stay extended over a region $\sim O(N^0)$. This is because the contribution to the free energy due to entropy is $ O(N)$ which is of the same order as the energy contribution. This is in sharp contrast to the HR [Sec.~\ref{hrd}] and HC [Sec.~\ref{hcm}] models as discussed in the previous sections where the length scale increases with $N$. Hence a field theory construction does not make sense for both the Toda model and the Harmonic chain due to the lack of a macroscopic length scale.

Therefore in order to understand the equilibrium properties of such models, we have performed detailed MC simulations of the Toda model with quadratic and quartic traps. In Fig.~\ref{fig:Toda} we plot the density profile for the Toda model at $T = 1$ for $d = 0.01, 1.0,  10.0$ with $N=32, 64,128$. For $d = 0.01, 1.0, 10.0$, we find that the density profiles converge with increasing system size and have $O(N^{0})$ spread as expected. However, for $d = 0.1$ we do not observe the convergence for the system sizes studied.

It is interesting to consider two special limits in the interaction length scale ($d$) in the Toda model. For small $d \ll 1$, the Toda interactions become similar to hard point gas, and for large $d \gg 1$ it can be approximated by nearest neighbor weakly interacting harmonic chain. In these two limits, the density profile has the form $\sim e^{-\beta U_{\delta}(x)}$. More precisely in our case we get
\begin{align}\label{toda:limit}
	\rho_T(x,\beta) \approx \frac{\beta^{\frac{1}{\delta}}}{\delta^{\frac{1}{\delta}} \Gamma\left[1+1/\delta \right]}{\rm exp}\left(-\beta \frac{x^{\delta}}{\delta}\right),	
\end{align}
for both quadratic $(\delta = 2)$ and quartic trap $(\delta = 4)$ as verified in Fig. \ref{fig:Toda_a_limits}. Furthermore, note that the free energy of the Toda model $\mathcal{F}_T(\beta, d, J)$ at temperature $1/\beta$ can be related to the free energy of a Toda model at $\beta = 1$ as
\begin{align}\label{toda:scaledfree}
 \mathcal{F}_T(\beta, d, J) = \mathcal{F}_T(1, d \beta^{\frac{1}{\delta}}, \beta J) + \frac{N}{\beta \delta} \ln \beta.
\end{align}
Interestingly, Eq.~\eqref{toda:scaledfree} implies that the Toda model at any temperature can be mapped to Toda model at temperature $1/\beta = 1$ with rescaled interaction strength $J \to \beta J$ and interaction length scale $d \to d \beta^{\frac{1}{\delta}}$. Therefore studying density profiles for various values of $d$ is equivalent to studying the density profiles for different values of temperatures. Note that, the Toda model at high temperatures ($\beta \ll 1$) can be approximated as a hard point gas ($d \ll 1$) and for low temperatures ($\beta \gg 1$) it can be approximated as a harmonic chain with nearest neighbor interactions ($d \gg 1$) under an external pressure.

To better understand the features of the density profiles of the Toda model we study its harmonic limit [Eq.~\eqref{eqn:harm} with $k_2 = 1/d^2$ and $k_3 = 1/d$],  which is analytically tractable for the quadratic trap. We obtain the exact expression (see Appendix~\ref{AppIII}) for the density profile of the harmonic chain in a quadratic trap which is given by (see Appendix~\ref{AppIII})
\begin{align}\label{eqn:hrmcdensity_main}
	\rho_H\left(x, \beta\right) 
	&= \frac{1}{N}\sum_{l=1}^N \frac{1}{\sqrt{2\pi {\rm Var}(x_l)}}{\rm exp}\left(-\frac{\Big(x-\langle x_l\rangle_{\beta}\Big)^2}{2~{\rm Var}(x_l)}\right),
\end{align}
where the mean position of the $l^{\rm th}$  particle $\langle x_l \rangle_{\beta}$ and its variance ${\rm Var}(x_l)$  are given in Eq.~\eqref{harm:kappal} and Eq.~\eqref{harm:varl} of Appendix~\ref{AppIII} respectively. This is plotted in Fig.~\ref{fig:harmvstoda} and compared with the corresponding MC simulation of the Toda model for $d = 1.0$ and $10.0$. We find good agreements both for $d = 1.0, 10.0$. Note that already for $N = 32, 64, 128$ the density profiles have converged to an $N$-independent form. As mentioned earlier for the Toda model at $d=0.1$, a slower convergence with $N$ was observed for the density profile (see Fig~\ref{fig:Toda} b, f). Interestingly a similar slow convergence can be analytically demonstrated (see Appendix~\ref{AppIII}) for a stiff (large $k_2$) harmonic chain. In Fig.~\ref{fig:harm2}, the density profiles for the harmonic chain with $T = 1$, $k_1=1$,$k_2 = 100$ and $k_3=10$ for $N = 32, 64, 128, 256$ and $N \to \infty$ are shown. It can be seen that for increasing $N$ the density profiles converges slowly to the $N\to \infty$ curve.
\begin{center}
	\begin{table}
		\begin{tabular}{|c|c|c|}
			\hline
			{Model}~\textbackslash~{Trap} & Harmonic ($\delta=2$) & Quartic ($\delta=4$)\\
			\hline
			Hard rods & Eq. \eqref{dens:hrd-scal},& Eq. \eqref{dens:hrd-scal},\\
			& $\alpha_R =1$, $\gamma_R =2$,& $\alpha_R =1$, $\gamma_R =4$, \\
			&  $\lambda_R = 2$&  $\lambda_R = 4$\\
			& Figs.~\ref{fig:HRods}a, \ref{fig:HRods}b, \ref{fig:HRods}c&  Figs.~\ref{fig:HRods}d, \ref{fig:HRods}e, \ref{fig:HRods}f\\
			\hline
			Hyperbolic Calogero & Eq. \eqref{dens:hcm-scal}, & Eq. \eqref{dens:hcm-scal},\\
			& $\alpha_C =\frac{1}{2}$, $\gamma_C =  1$, & $\alpha_C =\frac{1}{3}$, $\gamma_C =\frac{4}{3}$, \\
			& $\lambda_C =  1$ & $\lambda_C =\frac{4}{3}$\\
			& Figs.~\ref{fig:HCM}a, \ref{fig:HCM}b, \ref{fig:HCM}c&  Figs.~\ref{fig:HCM}d, \ref{fig:HCM}e, \ref{fig:HCM}f\\
			\hline
		\end{tabular}
		\caption{A summary of the scaling behavior of the densities for the hard rods (HR) and the hyperbolic Calogero (HC) model in quadratic and quartic traps.}\label{table1}
	\end{table}
\end{center}

\section{Conclusion}
\label{conclusion}
To summarize, we have presented the equilibrium density profiles at finite temperatures of two integrable models, the hard rods and the hyperbolic Calogero model, in quadratic and quartic traps. For these models inter-particle repulsion is strong enough to avoid particle trajectories from crossing. The trap confines these systems spatially and breaks integrability. For these two models, we studied equilibrium density profiles using a field theory approach and Monte-Carlo (MC) simulations.

We developed appropriate field theory for these two models by extending the approaches used in Ref. ~\cite{agarwal2019harmonically}. From  the field theory we computed the equilibrium density profiles, and their dependence on system size $N$ and temperature $T$. The field theory calculations predict precise scaling forms for the equilibrium density profiles with respect to $N$ and $T$. A summary of the scaling forms are given in Table~\ref{table1}. We find that the predictions from field theory for hard rods agree remarkably well with MC simulations (Fig.~\ref{fig:HRods}). For the hyperbolic Calogero model the agreement is also reasonably good (Fig.~\ref{fig:HCM}). On the other hand for integrable models that allow crossing of particle trajectories, such as the Toda model in quadratic and quartic trap, a field theoretic description is ill-suited due to the lack of a macroscopic length scale. For this case, we have presented microscopic analytical calculations, by employing Hessian approximation, and results from MC simulations.

Our work provides a framework for investigating the non-equilibrium dynamics, thermalization and transport in integrable models confined in external potentials. More precisely, one can ask whether these systems under Hamiltonian dynamics are ergodic, chaotic, and whether or not they equilibrate/thermalize, when placed in different confining traps. This is an area of active current research both theoretically~\cite{bulchandani2021quasiparticle} and experimentally~\cite{kinoshita2006quantum}.
%

\vspace{1cm}
\section*{Acknowledgements}
M.K. would like to acknowledge support from the project 6004-1 of the Indo-French Centre for the Promotion of Advanced Research (IFCPAR), Ramanujan Fellowship (SB/S2/RJN-114/2016), SERB Early Career Research Award (ECR/2018/002085) and SERB Matrics Grant (MTR/2019/001101) from the Science and Engineering Research Board (SERB), Department of Science and Technology (DST), Government of India. A.K. acknowledges the support of the core research grant  CRG/2021/002455 and the  MATRICS grant MTR/2021/000350 from the SERB, DST, Government of India. A.D., M.K., and A.K. acknowledges support of the Department of Atomic Energy, Government of India, under Project No. 19P1112R\&D.

\appendix
\vspace{1cm}
\section{Derivation of the free energy} \label{AppI}
To obtain the free energy $\mathcal{F}[\rho, \beta]$  in Eq.~\eqref{free:sys}, one first needs to compute $f_{\rm int}(\rho(x), \beta)$ defined in Eq.~\eqref{free:int-sub}, where we recall that $f_{\rm int}(\rho(x), \beta)$ is the contribution to the free energy of the subsystem (recall Fig.~\ref{fig:schematic}) due to interactions (i.e., excluding the external confining potential). In the following, we present the calculation of $f_{\rm int}(\rho(x), \beta)$ for the  hard rods (HR) model in Appendix~\ref{appendix:free_rod} and the hyperbolic Calogero (HC) model in Appendix~\ref{appendix:free_calo}  separately.

\subsection{Free energy for hard rods}
\label{appendix:free_rod}

The free energy per particle for hard rods of length $a$, $f_{\rm int}(\rho(x), \beta)$, can be calculated using the partition function [term in the parenthesis (square bracket) of Eq.~\eqref{part:subsystem}]
\begin{align}\label{hrd:partition_sub}
	\notag &Z_{\rm int}(n_s, x_s, \Delta, \beta)\\ \notag &= \int_{x_s-\Delta/2+\frac{a}{2}}^{x_s+\Delta/2-(n_s-\frac{1}{2})a} dy_1...\int^{x_s+\Delta/2-(n_s-i+\frac{1}{2})a}_{y_{i-1}+a} dy_i...\times \\
	&\quad\quad\quad\quad\quad\quad\quad\quad\quad\quad\quad\quad\quad\quad\quad\int^{x_s+\Delta/2-\frac{a}{2}}_{y_{n_s-1}+a} dy_{n_s}\notag\\
	&= \int_{\frac{a}{2}}^{\Delta-(n_s-\frac{1}{2})a} dy_1...\int^{\Delta-(n_s-i+\frac{1}{2})a}_{y_{i-1}+a} dy_i...
	\int^{\Delta-\frac{a}{2}}_{y_{n_s-1}+a} dy_{n_s},
\end{align}
where $y_i$ is the position of the $i^{\rm th}$ rod of the subsystem which is centered at $x_s$ and has a size $\Delta$. In each subsystem there are $n_s$ hard rods. Note that, since the integrand in the second line of Eq.~\eqref{hrd:partition_sub} is constant and translationally invariant, we have shifted the limits of the integrals from $y_i \to y_i-(x_s-\Delta/2)$. This shift results in the integrals given in the third line of Eq.~\eqref{hrd:partition_sub}. These integrals can be computed using the variable transformation $z_i = y_i - (i-\frac{1}{2})a$, which gives
\begin{align}
	Z_{\rm int}(n_s, x_s, \Delta, \beta) 
	&={\rm exp} \Bigg[n_s \log \left(\frac{1-\rho(x_s)a}{\rho(x_s)}\right) -n_s\Bigg],
\end{align}
where we introduce the density in the given subsystem 
\begin{align}
	\rho(x_s) = \frac{n_s}{\Delta}.
\end{align}
The free energy per particle in a given subsystem in the large $n_s$ limit is given by ~\cite{tonks1936complete}
\begin{align}\label{hrd:freeEnergyperparticle}
	f_{\rm int}\left(x_s, \beta\right) \notag &= -\frac{1}{n_s \beta}\log\Big[Z_{\rm int}(n_s, x_s, \Delta, \beta)\Big]\\
	&=-\frac{1}{\beta} \log\left[\frac{1-  a \rho(x_s)}{\rho(x_s)}\right] + \frac{1}{\beta}.
\end{align}
One can see that, from the partition function in Eq.~\eqref{hrd:partition_sub}, the logarithmic term in Eq.~\eqref{hrd:freeEnergyperparticle} is the configurational entropy which includes the effect of hard rod exclusion. Note that the free energy due to interaction is a function of the density field and we rewrite the arguments of 
\begin{align}
	f_{\rm int}(x_s, \beta) &\equiv f_{\rm int}(\rho(x_s), \beta).
\end{align}
The total (i.e. including the contribution due to the external potential) free energy of the entire system $\mathcal{F}_R[\rho(x_s), \beta]$ is obtained by summing over the total free energy
\begin{align}
	n_sf(x_s, \beta) = n_sf_{\rm int}(\rho(x_s),\beta) + n_sU_{\delta}(x_s),
\end{align}
associated with each subsystem. We therefore get 
\begin{align}
	\mathcal{F}_R[\rho(x_s), \beta]&=\sum_{s=1}^{N_b} n_sf(x_s, \beta)\notag,\\
	& = \sum_{s=1}^{N_b}~\rho(x_s) ~\Delta~\Big[f_{\rm int}(\rho(x_s), \beta) + U_{\delta}(x_s)\Big].
\end{align}
Converting the summation to integration i.e., $\sum_{s=1}^{N_b} \Delta \to \int_{-\infty}^{\infty} dx$ we obtain
\begin{align}\label{hrd:freerhos} 
	\mathcal{F}_R[\rho(x), \beta]& = \int_{-\infty}^{\infty} dx~ \rho(x) \Big[f_{\rm int}(\rho(x), \beta) + U_{\delta}(x)\Big]
\end{align}
Using Eq.~\eqref{hrd:freeEnergyperparticle} in Eq.~\eqref{hrd:freerhos}, we obtain
\begin{align}\label{hrd:finalfree} 
	\mathcal{F}_R \left[\rho(x), \beta\right] = \int_{-\infty}^{\infty}dx~\rho(x) \Bigg[ U_{\delta}(x) -\frac{1}{\beta} \log\left(\frac{1-  a \rho(x)}{\rho(x)}\right) \Bigg],
\end{align}
which is the free energy of the hard rods in an external trap $U_{\delta}(x)$ given in Eq.~\eqref{free:hrd-sys1} of the main text. In Eq.~\eqref{hrd:finalfree} we have ignored the density independent term from Eq.~\eqref{hrd:freeEnergyperparticle}.

\subsection{Free energy for hyperbolic Calogero model}
\label{appendix:free_calo}

The field theoretic description of the hyperbolic Calogero model in external traps is well understood~\cite{gon2019duality}. In this section we present an alternative derivation of the total free energy $\mathcal{F}_C[\rho(x), \beta]$ of the system. Using the approximate scheme described in Refs.~\cite{stone2008classical, agarwal2019harmonically} we compute the free energy per particle $f_{\rm int}(\rho(x), \beta)$ of the subsystem due to the interaction which is described below. 

The free energy per particle for the HC model, $f_{\rm int}(\rho(x), \beta)$, can be calculated using the partition function which we recall to be
\begin{align}\label{part:subsystem_app}
	Z_{\beta}(n_s, x_s, \Delta) &\approx \notag {\rm exp} \big(-\beta n_s U_{\delta}(x_s)\big)\times\\ & \Bigg[\int_{x_s-\frac{\Delta}{2}}^{x_s+\frac{\Delta}{2}} {\bf dx}_{n_s} \prod_{\substack{i, j=1 \\ j\neq i}}^{n_s} {\rm exp}\Big(-\frac{\beta}{2} \left[ V(x_i-x_j)\right]\Big)\Bigg] .
\end{align}
For the HC model Eq.~\eqref{part:subsystem_app} becomes
\begin{align}
	Z_{\beta}(n_s, x_s, \Delta) &\approx \notag {\rm exp} \big(-\beta n_s U_{\delta}(x_s)\big) Z_{\rm int}(n_s,x_s, \Delta, \beta),
\end{align} 
where
\begin{align}\label{hcm:partition_sub}
	Z_{\rm int}(n_s,x_s, \Delta, \beta) = \int_{0}^{\Delta}& {\bf dx}_{n_s} \times \notag \\ &{\rm exp} \left[-\frac{\beta J}{2} \sum_{i=1}^{n_s} \sum_{\substack{j=1 \\ j\neq i}}^{n_s} \frac{1}{\sinh^2\left(|x_i-x_j|\right)}\right],
\end{align}
where $x_i$ is the position of the $i^{\rm th}$ particle and $J$ is the interaction strength. As mentioned in the main text [Sec.~\ref{ftf}], the $x_{i}$ is a running integration variable not to be confused with the position of the center of the subsystem $x_s$. One can approximate the exponential term in the integrand of Eq.~\eqref{hcm:partition_sub} as
\begin{align}
	&{\rm exp} \left(-\frac{\beta J}{2} \sum_{i=1}^{n_s} \sum_{\substack{j=1 \\ j\neq i}}^{n_s} \frac{1}{\sinh^2\left(|x_i-x_j|\right)}\right) \notag \\&\notag  \approx {\rm exp} \left(-\frac{\beta J}{2} \frac{n_s^2}{\Delta^2} \sum_{i=1}^{n_s} \sum_{\substack{j=1 \\ j\neq i}}^{n_s} \frac{1}{\left(|i-j|\right)^2}\right),\\
	& \label{append:hcm:eq2}\approx{\rm exp} \left(-\beta J \frac{n_s^3}{\Delta^3}\zeta(2)\Delta\right),
\end{align}
where in the second line of Eq.~\eqref{append:hcm:eq2} we approximated $x_i \approx i\Delta/n_s$ for all $i$ since $\Delta$ is assumed to be small enough to ensure uniform density over the subsystem. We have also used $\zeta(2) = \sum_{i=1}^{\infty} 1/i^2$, where $\zeta(k)= \sum_{i=1}^{\infty} 1/i^k$ represents the Riemann zeta function. Using Eq.~\eqref{append:hcm:eq2} the partition function in Eq.~\eqref{hcm:partition_sub} takes the form
\begin{align}
	Z_{\rm int}(n_s,x_s, \Delta, \beta) &\approx {\rm exp} \left(-\beta J \frac{n_s^3}{\Delta^3}\zeta(2)\Delta\right) \int_{0}^{\Delta}{\bf dx}_{n_s},\notag \\
	&={\rm exp} \left(-\beta J \frac{n_s^3}{\Delta^3}\zeta(2)\Delta\right)\frac{\Delta^{n_s}}{n_s!},
\end{align}
which can be rewritten using Stirling's approximation $\log[n!] \approx n~\log [n] - n$ as
\begin{align}\label{hcm:part3}
	Z_{\rm int}&(n_s,x_s, \Delta, \beta) \notag\\&\asymp {\rm exp} \Bigg(-n_s\Big[\log\big[\rho(x_s)\big]+\zeta(2)\beta J \rho(x_s)^2\Big]\Bigg),
\end{align}
where we recall that $\rho(x_s) = n_s/\Delta$. Hence, using the first line in Eq.~\eqref{hrd:freeEnergyperparticle}, the free energy per particle of the subsystem becomes
\begin{align}\label{hcm:free-a1}
	f_{\rm int} \left(\rho(x_s), \beta\right) = J \zeta(2) \rho(x_s)^2+\frac{1}{\beta} \log\left[\rho(x_s)\right].
\end{align}
Similar to procedure detailed in Appendix~\ref{appendix:free_rod}, using the above expression Eq.~\eqref{hcm:free-a1}  we can compute the total free energy of the system as
\begin{align}
	\mathcal{F}_C \left[\rho(x), \beta\right] = \int_{-\infty}^{\infty}~dx~\rho(x)& \Bigg( U_{\delta}(x) +J \zeta(2) \rho(x)^2 \notag \\ &+\frac{1}{\beta} \log\rho(x)\Bigg),
\end{align}
which is the expression for the free energy [see Eq.~\eqref{hcm:free_rho} of the main text] of the HC model in an external trap $U_{\delta}(x)$.

\section{Analytical forms of density profiles for hard rods and hyperbolic Calogero model at low and high rescaled temperatures~$c$} \label{AppII}
To obtain the exact analytical expression for the equilibrium density profiles by solving the transcendental equations, Eq.~\eqref{hrd:s-chem} for HR model and Eq.~\eqref{hcm:scaling-chem2} for HC model, is a highly non-trivial. However, we can obtain approximate expressions for the densities at low ($c \ll 1$) and high ($c \gg 1$) rescaled temperatures $c$. For $c \ll 1$, this is done by approximating the equilibrium density profile to be a small deviation around the zero temperature density profile. On the other hand, for $c \gg 1$, the particles expected to spread far apart, thereby diluting the gas. Thus in this regime ($c \gg 1$) it is reasonable to assume the density to be very small. In this section, using the above mentioned assumptions for the saddle point equations i.e., Eq.~\eqref{hrd:s-chem} for HR model and Eq.~\eqref{hcm:scaling-chem2} for HC model, we discuss the analytical forms of the density profiles.

\subsection{Analytical forms of the density profiles for hard rods}
\label{appendix:hardrod}
In this subsection, we discuss the case of hard rods, recall that the saddle point equation for the hard rods is 
\begin{align}\label{hrd:s-chem_app}
	\mu_R(c) = \frac{y^{\delta}}{\delta} -c \Bigg[\log\left(\frac{1-a~\rho_R (y,c)}{\rho_R (y,c)}\right) - \frac{1}{1-a~\rho_R (y,c)}\Bigg].
\end{align}
We now analyze Eq.~\eqref{hrd:s-chem_app} for both small and large rescaled temperatures $c$. In the following, we use the value of chemical potential $\mu_R(c)$ which is obtained by numerical solving Eq.~\eqref{hrd:s-chem_app} with the constraint that the density is normalized to unity.

\subsubsection{Small rescaled temperature $c \ll 1$}
\label{appendix:hrd_low_c}
At zero temperature i.e., $c =0$, all the hard rods arrange themselves leaving no gaps. In other words the center to center distance between the rods is $a$, thereby making the density $\rho_N(x, 0) = N/a$ where we recall that $N$ is number of hard rods. In the rescaled density variables this corresponds to a scaled density profile 
\begin{align}\label{hrd:zerotemp}
	\rho_R(y, 0) = 
	\begin{cases}
		\frac{1}{a}&~\text{for}~ y\leq \left|\frac{a}{2}\right|\\
		0 &~\text{for}~ y>\left|\frac{a}{2}\right|.
	\end{cases}
\end{align}
We now study the effects of turning on a small temperature. More precisely we address how the zero temperature profile given in Eq.~\eqref{hrd:zerotemp} gets smeared. Note that at
\begin{align}\label{hrd:yc}
	y = y_c = \left(\mu_c \delta\right)^{\frac{1}{\delta}}.
\end{align}
the square bracket in Eq.~\eqref{hrd:s-chem_app} changes sign. This in turn determines the following three distinct regions 
\begin{enumerate}
	\item[(i)] Bulk region ($|y|<y_c$): The density profile deviates from the value $1/a$.
	\item[(ii)] Edge region (a zone where $|y-y_c| \lesssim O(c)$): The density profile deviates from a value $\rho_R(y_c, c) = \rho^*_R$ which is the density at $y = y_c$. At this value of $\rho^*_R$ the square bracket in Eq.~\eqref{hrd:s-chem_app} becomes zero.
	\item[(iii)] Tail region ($|y|>y_c$): The density profile for finite temperature in this region deviates from its zero temperature value $\rho_R(y, 0)=0$.
\end{enumerate}
We now compute the density profile at low temperatures of these three regions separately. 
\vskip 0.5cm
\textit{(i) Bulk region $|y|<y_c$} : In this region, we assume that the density takes the form 
\begin{align}
	\label{hrd:rhocorr}
	\rho_R(y,c) = \frac{1}{a} + \rho_{1}(y,c),
\end{align}
where $\rho_{1}(y,c)$ denotes the deviation about the zero temperature density. For sake of brevity, we henceforth omit the arguments of $\rho_{1}(y,c)$. Using Eq.~\eqref{hrd:rhocorr} in Eq.~\eqref{hrd:s-chem_app} we get
\begin{align}\label{hrd:s-chem_app2}
	\mu_c -\frac{y^{\delta}}{\delta} &=  - c\Bigg[\log{\left(\frac{-a^2\rho_1}{1+a~\rho_1}\right)} +\frac{1}{a~\rho_1}\Bigg].
\end{align}
It turns out that a convenient perturbation parameter is the following
\begin{align}\label{hrd:nu}
	\nu(y) = \frac{c\delta}{\left(y_c^{\delta}-y^{\delta}\right)},
\end{align}
where we have used Eq.~\eqref{hrd:yc}. 
Using Eq.~\eqref{hrd:nu} in Eq.~\eqref{hrd:s-chem_app2} we obtain
\begin{align}\label{hrd:s-chem_app3}
	-\frac{1}{\nu(y)} &=\log{\left(\frac{-a^2\rho_1}{1+a~\rho_1}\right)} +\frac{1}{a~\rho_1}.
\end{align}
To solve Eq.~\eqref{hrd:s-chem_app3} we first perform a Taylor expansion
\begin{align}
	\label{hrd:rho1}
	-\frac{1}{\nu(y)} &= \log{\left(-a^2\rho_1\right)} -a~\rho_1-\frac{a^2}{2}\rho_1^2 +\frac{1}{a~\rho_1},
\end{align}
which can be again rearranged to give
\begin{align} 
	\label{hrd:rho1a}
	a~\rho_1 &=-\nu(y)\frac{1}{1+ \nu(y)\left(\log{\left[-a^2\rho_1\right]} -a~\rho_1-\frac{a^2}{2}\rho_1^2\right)}.
\end{align}
We perform a Taylor series expansion [upto second order in $\nu(y)$] of the fraction on the right hand side of Eq.~\eqref{hrd:rho1a}, since $\nu(y) \ll 1$. This gives
\begin{align}
	\label{hrd:rho11}
	\rho_1 &\approx -\frac{\nu(y)}{a}
	\Bigg[1-\nu(y)\left(\log{\left[-a^2\rho_1\right]} -a~\rho_1-\frac{a^2}{2}\rho_1^2\right) \notag \\&+\nu(y)^2\left(\log{\left[-a^2\rho_1\right]} -a~\rho_1-\frac{a^2}{2}\rho_1^2\right)^2\Bigg].
\end{align}
Since the correction to the zero temperature density $\rho_1 \ll 1$ and $\nu(y) \ll 1$, we can invert Eq.~\eqref{hrd:rho11} to express $\rho_1$ as a function of $\nu(y)$ order by order. This gives
\begin{align}
	\rho_1^{(0)} &= -\frac{\nu(y)}{a},\label{hrd:lowcdens0th}\\
	\rho_1^{(1)} &= -\frac{\nu(y)}{a}+\frac{\nu(y)^2}{a}\log \Big[a\nu(y)\Big],\label{hrd:lowcdens1st}\\
	\rho_1^{(2)} &= -\frac{\nu(y)}{a}+\frac{\nu(y)^2}{a}\log\Big[ a\nu(y)\Big]-\frac{\nu(y)^3}{a}\log \Big[a\nu(y)\Big]^2,\label{hrd:lowcdens2nd}
\end{align}
where recall that $\nu(y)$ is defined in Eq.~\eqref{hrd:nu}. The superscript associated with $\rho_1$ in  Eqs.~\eqref{hrd:lowcdens0th}-\eqref{hrd:lowcdens2nd} represents their respective orders. In Figs.~\ref{fig:hrod_dens_low}a and \ref{fig:hrod_dens_low}d using Eqs.~\eqref{hrd:lowcdens0th}-\eqref{hrd:lowcdens2nd}, we find a good agreement between the analytically obtained series solutions and the numerical solution of Eq.~\eqref{hrd:s-chem_app}.

\textit{(ii) Edge  region $|y-yc| \lesssim O(c)$:}  Recall that this region is a zone defined by $|y-yc| \lesssim O(c)$. Here $\nu(y) \gtrsim O(1)$, and therefore, the above expressions Eqs.~\eqref{hrd:lowcdens0th}- \eqref{hrd:lowcdens2nd} fail. Hence, in this zone [$|y_c - y| \lesssim O(c)$], we assume that the density takes the form
\begin{align}
	\rho_R(y,c) = \rho_R^* + \phi(y),
\end{align}
where $\rho_R^*$ is the value of the density at $y = y_c$ and the correction $\phi(y) \ll 1$. The value of $\rho_R^*$ can be obtained by numerically solving  Eq.~\eqref{hrd:s-chem_app} at $y = y_c$ which gives
\begin{align}\label{hrd:s-chemy=yc}
	\log{\left(\frac{1-a \rho_R^*}{\rho_R^*}\right)} - \frac{1}{1-a\rho_R^*} = 0.
\end{align}
In this region, we define the perturbation parameter 
\begin{align}\label{hrd:by}
	b(y) = 1-\exp\left[-\frac{y_c^{\delta}-y^{\delta}}{c \delta}\right] \ll 1.
\end{align}
Using Eq.~\eqref{hrd:by} in Eq.~\eqref{hrd:s-chem_app} we get
\begin{align}
	\log{\left[1-b(y)\right]} &=  \log{\Bigg[\frac{1-a \rho_R^* -a \phi(y)}{\rho_R^* + \phi(y)}\Bigg]} \notag \\ &- \frac{1}{1-a\rho_R^*-a\phi(y)},
\end{align}
which upon Taylor series expansion, upto third order in $\phi(y)$, yields
\begin{align}\label{hcm:edge-b}
	b(y) &\approx \phi(y) \left(\frac{1}{\rho_R^*\left(1-a\rho_R^*\right)^2}\right) \notag \\ &- \phi(y)^2\left(\frac{-1 + 3 a \rho_R^*}{2 \rho_R^{*2} (1-a\rho_R^*)^3 }\right)\notag \\ &- \phi(y)^3\left(\frac{-1 + 4 a \rho_R^*-6 (a \rho_R^*)^2}{3 \rho_R^{*3} (1-a\rho_R^*)^4 }\right).
\end{align}
We can represent the correction to density $\phi(y)$ as a function of $b(y)$ by inverting Eq.~\eqref{hcm:edge-b} order by order which gives 
\begin{align}
	\phi^{(0)}(y) &= b(y)\rho_R^*(1-a\rho_R^*)^2,\label{hrd:lowedge0}\\
	\phi^{(1)}(y) &= b(y)\rho_R^*(1-a\rho_R^*)^2 \notag \\& +  \frac{b(y)^2}{2}\rho_R^*(1-a\rho_R^*)^3\big(-1+3a \rho_R^*\big)\label{hrd:lowedge1},\\
	\phi^{(2)}(y) &= b(y)\rho_R^*(1-a\rho_R^*)^2 \notag \\& + \frac{b(y)^2}{2}\rho_R^*(1-a\rho_R^*)^3\big(-1+3a \rho_R^*\big)\notag \\& + \frac{b(y)^3}{6}\rho_R^*(1-a\rho_R^*)^4\Big(1-10a \rho_R^*+15 (a\rho_R^*)^2\Big)\label{hrd:lowedge2}.
\end{align}
In Figs.~\ref{fig:hrod_dens_low}b and \ref{fig:hrod_dens_low}e, using Eqs.~\eqref{hrd:lowedge0}-\eqref{hrd:lowedge2}, we compare the analytically obtained series solutions with the numerical solution of Eq.~\eqref{hrd:s-chem_app} and see reasonable agreement.

\textit{(iii) Tail region $|y|>y_c$:} In this region we assume that the density is very small and takes the form $\rho_R(y,c) = \rho_1$ with $\rho_1 \ll 1$. Using this assumption in Eq.~\eqref{hrd:s-chem_app} we get
\begin{align}\label{hrd:s-chem2}
	\mu_c &= \frac{y^{\delta}}{\delta} - c\Bigg[\log{\left(\frac{1-a~ \rho_1}{\rho_1}\right)} - \frac{1}{1-a~\rho_1}\Bigg].
\end{align}
We introduce the perturbation parameter
\begin{align} \label{hrd:epsilony}
	\epsilon(y) = \exp\left(\frac{y_c^{\delta} -y^{\delta}}{c \delta}\right).
\end{align}
Since in this region $|y| > y_c$ and $c \ll 1$, it implies $\epsilon(y) \ll 1$. Using Eq.~\eqref{hrd:epsilony} in Eq.~\eqref{hrd:s-chem_app} we get
\begin{align}
	\log{\big[\epsilon(y)\big]} \label{hrd:s-chem4} &= -\log{\left(\frac{1-a~ \rho_1}{\rho_1}\right)} + \frac{1}{1-a~\rho_1}.
\end{align}
After some algebra and assuming $\rho_1 \ll 1$ in Eq.~\eqref{hrd:s-chem4}, we obtain the transcendental equation
\begin{align}
	\rho_1 &\approx \frac{\epsilon(y)}{e} \exp\Bigg(-2 a \rho_1- \frac{(a \rho_1)^2}{2} \Bigg),
\end{align}
where $e\approx2.71828$ is the Euler's number. We can now represent the density in terms of $\epsilon(y)/e$ as a series given by
\begin{align}
	\rho_1^{(0)} &= \frac{\epsilon(y)}{e},\label{hrd:lowtail0}\\
	\rho_1^{(1)} &= \frac{\epsilon(y)}{e} - 2 a \left(\frac{\epsilon(y)}{e}\right)^2,\label{hrd:lowtail1}\\
	\rho_1^{(2)} &= \frac{\epsilon(y)}{e} - 2 a \left(\frac{\epsilon(y)}{e}\right)^2+ \frac{11}{2}a^2 \left(\frac{\epsilon(y)}{e}\right)^3\label{hrd:lowtail2}.
\end{align}
In Figs.~\ref{fig:hrod_dens_low}c and~\ref{fig:hrod_dens_low}f, using Eqs.~\eqref{hrd:lowtail0}-\eqref{hrd:lowtail2}, we show a good agreement between the analytically obtained series solutions with the numerical solution of Eq.~\eqref{hrd:s-chem_app}. Recall that the chemical potential $\mu_R(c)$ in the perturbation parameter $\epsilon(y)$ given in Eq.~\eqref{hrd:epsilony} is obtained from the numerical solution of Eq.~\eqref{hrd:s-chem_app} along with the constraint of unit normalized density.

\subsubsection{Large rescaled temperatures: $c \gg 1$}
\label{appendix:hrd_high_c}
When the temperature is high the particles explore a larger region in space thereby diluting the system as a consequence of which we get $\rho_R(y,c) \ll 1$. We introduce a convenient perturbation parameter
\begin{align}
	\eta(y) = \exp\left(\frac{\mu_c\delta - y^{\delta}}{c\delta}\right),
\end{align}
where $\eta(y) \ll 1$, since the chemical potential  [see Fig.~\ref{fig:rod_mu} in the main text], obtained by numerically solving Eq.~\eqref{hrd:s-chem_app}, is negative and diverges for $c \gg 1$. 

We use the approximation $\rho_R(y,c) \ll 1$ in Eq.~\eqref{hrd:s-chem_app}, and use a procedure mathematically similar to the low temperature tail region (Appendix.~\ref{appendix:hrd_low_c}), to obtain the expressions
\begin{align}
	\rho_R^{(0)}(y,c) &= \frac{\eta(y)}{e},\label{hrd:high0}\\
	\rho_R^{(1)}(y,c) &= \frac{\eta(y)}{e} - 2 a \left(\frac{\eta(y)}{e}\right)^2,\label{hrd:high1}\\
	\rho_R^{(2)}(y,c) &= \frac{\eta(y)}{e} - 2 a \left(\frac{\eta(y)}{e}\right)^2+ \frac{11}{2}a^2 \left(\frac{\eta(y)}{e}\right)^3\label{hrd:high2}.
\end{align}
Note that the superscript in Eqs.~\eqref{hrd:high0}-\eqref{hrd:high2} represents the order in $\eta(y)$. In Fig.~\ref{fig:hrod_dens_high}, we see a decent agreement of the analytically obtained series solutions with the numerical solution of Eq.~\eqref{hrd:s-chem_app}.

\subsection{Asymptotic densities for hyperbolic Calogero model}
\label{appendix:low_c_C}
In the following subsections, we compute the asymptotic densities for the hyperbolic Calogero model at low and high rescaled temperature $c$ using a similar approach as described above for the HR model in Appendix~\ref{appendix:hardrod}. Here we recall that the saddle point equation for the HC model is
\begin{align}\label{hcm:scaling-chem2_app}
	\mu_C(c) = \frac{y^{\delta}}{\delta} + 3 \zeta(2)\rho_C(y,c)^2 + c \log\rho_C(y,c).
\end{align}
As before we analyze Eq.~\eqref{hcm:scaling-chem2_app} for small and large $c$.  In the following, we use the value of chemical potential $\mu_C(c)$ which is obtained by numerically solving Eq.~\eqref{hcm:scaling-chem2_app} with the constraint that the density is normalized to unity.

\subsubsection{Small rescaled temperatures $c \ll 1$}
\label{appendix:hcm_low_c}
For $c \ll 1$ the densities $\rho_C(y, c)$ are assumed to be a small deviation from the zero temperature which is obtained by solving Eq.~\eqref{hcm:scaling-chem2_app} for $c = 0$. The density profile is then given by
\begin{align}
	\rho_C(y, 0) = 
	\begin{cases}
		A_{\delta}\left(l^{\delta}-y^{\delta}\right)^{\frac{1}{2}} & \text{for} \quad |y|<l\\
		0 & \text{for} \quad |y|>l,
	\end{cases}
\end{align}
where 
\begin{align}
	A_{\delta} = \left(3 \delta \zeta(2)\right)^{-\frac{1}{2}}
\end{align}
and the edge of the support of the density is given by
\begin{align}
	l = \left(\frac{\delta}{2 A_{\delta} {\rm Beta}\left(\frac{1}{\delta}, \frac{3}{2}\right)}\right)^{\frac{2}{2+\delta}}.
\end{align}
Here
\begin{align}
	{\rm Beta}(x, y) = \int_{0}^1~dr~r^{x-1}(1-r)^{y-1},
\end{align} 
is the Beta function. The zero temperature chemical potential is given by [Eq.~\eqref{hcm:muc0} of main text]
\begin{align}
	\mu_C(0) &= \left(\frac{\pi }{2}\right)^{\frac{\delta }{\delta +2}} \left(\frac{\delta ^{-1/\delta } \Gamma \left(\frac{3}{2}+\frac{1}{\delta }\right)}{\Gamma \left(1+\frac{1}{\delta }\right)}\right)^{\frac{2 \delta }{\delta +2}}.
\end{align}
Similar to the HR model [Appendix~\ref{appendix:hrd_low_c}], at low temperatures the value of chemical potential $\mu_C(c)$ [in Eq.~\eqref{hcm:scaling-chem2_app}],
determines the (i) bulk $|y|<y_c$, (ii) edge $|y-y_c| \lesssim O(c)$ and the (iii) tail regions $|y|> y_c$, where
\begin{align}\label{hcm:yc}
	y_c = \left( \mu_C(c) \delta\right)^{\frac{1}{\delta}}.
\end{align}
We compute the density profile at low temperatures separately for the three regions.\\
\indent 
\textit{(i) Bulk region $|y|<y_c$:} In this region we assume that the density takes the form 
\begin{align}\label{appendix:hcm:densmallc}
	\rho_C(y,c) = \rho_C(y, 0) + \rho_1(y,c),
\end{align}
where $\rho_1(y,c)$ is the correction to the zero temperature density. We use the following notations
in the rest of the calculations
\begin{align}\label{hcm:notation}
	\rho_1(y,c) \equiv \rho_1,&\quad \quad \rho_C(y,0) \equiv \rho_0 \notag\\
	\mu_C(c) \equiv \mu_c,&\quad \quad \mu_C(0) \equiv \mu_0.
\end{align}
Using Eq.~\eqref{hcm:notation} in Eq.~\eqref{hcm:scaling-chem2_app} gives
\begin{align}\label{appendix:hcm:muc}
	\mu_c &= \frac{y^{\delta}}{\delta} + c\log\rho_0+c\log\left[1+\frac{\rho_1}{\rho_0}\right]\notag \\&+3\zeta(2)\left(\rho_0+\rho_1\right)^2.
\end{align}
Since the temperature is low ($c \ll 1$), the correction to the zero temperature density $\rho_1 \ll \rho_0$. Furthermore we introduce the perturbation parameter in the bulk region
\begin{align}\label{hcm:nu}
	\nu(y) = \frac{\mu_0+c\log\rho_0-\mu_c}{c+6\zeta(2)\rho_0^2}\ll 1.
\end{align}
For $c \ll 1$, it turns out that $\mu_c$ and $\mu_0$ are very close, which implies $\nu(y) \ll 1$ and therefore a suitable perturbative parameter. Using Eq.~\eqref{hcm:nu} in Eq.~\eqref{appendix:hcm:muc} and expanding Logarithmic term up to $(\rho_1/\rho_0)^3$ gives 
\begin{align} \label{appendix:hcm:rho} 
	\frac{\rho_1}{\rho_0} &\approx -\nu(y)-\frac{1}{2}\left(\frac{\rho_1}{\rho_0}\right)^2-\frac{1}{3}\left(\frac{\rho_1}{\rho_0}\right)^3\frac{c}{c+6\zeta(2)\rho_0^2}.
\end{align}
We can solve Eq.~\eqref{appendix:hcm:rho} iteratively which gives
\begin{align}
	\frac{\rho_1^{(0)}}{\rho_0} &=-\nu(y),\label{hcm:lowcdens0th}\\
	\frac{\rho_1^{(1)}}{\rho_0} &= -\nu(y) -\frac{\nu(y)^2}{2},\label{hcm:lowcdens1st}\\
	\frac{\rho_1^{(2)}}{\rho_0} &= -\nu(y) - \frac{\nu(y)^2}{2}-\nu(y)^3\left(\frac{1}{2} - \frac{1}{3}\frac{c}{c+6\zeta(2)\rho_0^2}\right),\label{hcm:lowcdens2nd}
\end{align}
where the superscript represents their respective orders. In Fig.~\ref{fig:hhcm_dens_low}a and \ref{fig:hhcm_dens_low}d using the Eqs.~\eqref{hcm:lowcdens0th}-\eqref{hcm:lowcdens2nd} we find a good agreement between the analytically obtained series solution and the numerical solution of Eq.~\eqref{hcm:scaling-chem2_app}.

\textit{(ii) Edge region $|y-y_c| \lesssim O(c)$:}
Recall that this region is a zone defined by $|y-y_c| \lesssim O(c)$. Here $\nu(y)$ defined in Eq.~\eqref{hcm:nu} is no longer a small parameter and therefore ill-suited to be a perturbation parameter. Hence as in the case of HR model we assume the density to takes the form
\begin{align}\label{appendix:hcm:densedge}
	\rho_C(y,c) = \rho_C^* + \phi(y),
\end{align}
where $\rho_C^*$ is the value of the density at $y = y_c$ and the deviation $\phi(y) \ll \rho_C^*$. The value of $\rho_C^*$ can be obtained by numerically solving  Eq.~\eqref{hcm:scaling-chem2_app} at $y = y_c$ where $y_c$ is given in Eq.~\eqref{hcm:yc}. This gives 
\begin{align}\label{hcm:s-chemy=yc}
	3 \zeta(2) \rho_C^{*2} +c \log\rho_C^* = 0.
\end{align}
We now introduce a perturbative parameter 
\begin{align}\label{hcm:by}
	b(y) = \frac{y_c^{\delta}-y^{\delta}}{\delta\left(c + 6\zeta(2) \rho_C^{*2}\right)} \ll 1,
\end{align} 
since in this region $|y-y_c| \lesssim O(c)$ and $c \ll 1$. Substituting Eq.~\eqref{appendix:hcm:densedge} and using Eq.~\eqref{hcm:by} in Eq.~\eqref{hcm:scaling-chem2_app} and expanding, we get
\begin{align}
	\frac{\phi(y)}{\rho_C^*} &=  b(y) -\frac{1}{2}\left(\frac{\phi(y)}{\rho_C^*}\right)^2-\frac{1}{3}\left(\frac{\phi(y)}{\rho_C^*}\right)^3\frac{c}{c + 6\zeta(2)\rho^{*2}}.
\end{align}
Using a similar iterative approach as before we can represent the correction to the zero temperature density $\phi(y)$ as
\begin{align}
	\frac{\phi^{(0)}(y)}{\rho_C^*} &=b(y),\label{hcm:lowcdens0th1}\\
	\frac{\phi^{(1)}(y)}{\rho_C^*} &= b(y) - \frac{b(y)^2}{2},\label{hcm:lowcdens1st1}\\
	\frac{\phi^{(2)}(y)}{\rho_C^*} &= b(y) - \frac{b(y)^2}{2}+b(y)^3\left(\frac{1}{2} - \frac{1}{3}\frac{c}{c+6\zeta(2)\rho_0^2}\right).\label{hcm:lowcdens2nd1}
\end{align}
In Figs.~\ref{fig:hhcm_dens_low}b. and~\ref{fig:hhcm_dens_low}e, using Eqs.~\eqref{hcm:lowcdens0th1}-\eqref{hcm:lowcdens2nd1}, we find a good agreement between the analytically obtained series solution and the numerical solution of Eq.~\eqref{hcm:scaling-chem2_app}.

\textit{(iii) Tail region $|y| > y_c$:}
Unlike the bulk and the edge regions, in the tail region we assume that the density is small and takes the form $\rho_C(y,c) = \rho_1$ where $\rho_1 \ll 1$. Using this assumption in Eq.~\eqref{hcm:scaling-chem2_app} we obtain
\begin{align}\label{hcm:s-chem3}
	\mu_c &= \frac{y^{\delta}}{\delta} + c\log\rho_1 + 3 \zeta(2)\rho_1^2.
\end{align}
We now introduce a suitable perturbation parameter
\begin{align}\label{hcm:epsilony}
	\epsilon(y) = \exp\left(\frac{y_c^{\delta}-y^{\delta}}{c\delta}\right).
\end{align}
In the tail region, since $|y| > y_c$ and $c \ll 1$, the perturbation parameter $\epsilon(y) \ll 1$. Using Eq.~\eqref{hcm:epsilony} in Eq.~\eqref{hcm:s-chem3} and rearranging the terms gives
\begin{align} \label{appendix:hcm:densout}
	\rho_1 &= \epsilon(y)\exp\Bigg(-\frac{3\zeta(2)\rho_1^2}{c}\Bigg).
\end{align}
We can now represent the density in terms of $\epsilon(y)$ by using the iterative scheme, similar to bulk and edge regions, on Eq.~\eqref{appendix:hcm:densout}. This then gives
\begin{align}
	\rho_1^{(0)} = \epsilon(y)\label{hcm:tail0}&,\\
	\rho_1^{(1)} = \epsilon(y)\label{hcm:tail1}&\left[1 - \frac{3 \zeta(2)}{c}\epsilon(y)^{2}\right],\\
	\rho_1^{(2)} = \epsilon(y)\label{hcm:tail2}&\Bigg[1 - \frac{3 \zeta(2)}{c}\epsilon(y)^{2}+\frac{5}{2}\left(\frac{3 \zeta(2)}{c}\right)^2\epsilon(y)^{4}\Bigg].
\end{align}
In Fig.~\ref{fig:hhcm_dens_low}c and \ref{fig:hhcm_dens_low}f, using Eqs.~\eqref{hcm:tail0}-\eqref{hcm:tail2}, we show the analytically obtained asymptotic densities of the matches well with the numerical solution of Eq.~\eqref{hcm:scaling-chem2_app}.

\subsubsection{Large rescaled temperature: $c \gg 1$}
\label{appendix:hcm_high_c}
Similar to the HR model (Appendix~\ref{appendix:hrd_high_c}), when the temperature is high the particles are spread out over a larger region of space hence diluting the system as a consequence of which we get $\rho_C(y,c) \ll 1$. We again introduce a convenient perturbation parameter
\begin{align}\label{hcm:etay}
	\eta(y)= \exp\left(\frac{\mu_c\delta - y^{\delta}}{\delta c}\right).
\end{align}
Since the chemical potential (see Fig.~\ref{fig:hcm_mu} in the main text), obtained by numerically solving Eq.~\eqref{hcm:scaling-chem2_app}, is negative and diverges for $c\gg 1$, the perturbation parameter becomes small i.e., $\eta(y) \ll 1$. Using Eq.~\eqref{hcm:etay} along with  the low density approximation in Eq.~\eqref{hcm:scaling-chem2_app} and following a procedure mathematically similar to low temperature tail region (Appendix~\ref{appendix:hcm_low_c}) we obtain the expressions
\begin{align}
	\rho_C^{(0)}(y,c) = \eta(y)&,\label{hcm:high0}\\
	\rho_C^{(1)}(y,c) = \eta(y)&\left[1 - \frac{3 \zeta(2)}{c}\eta(y)^{2}\right],\label{hcm:high1}\\
	\rho_C^{(2)}(y,c) = \eta(y)&\Bigg[1 - \frac{3 \zeta(2)}{c}\eta(y)^{2}+\frac{5}{2}\left(\frac{3 \zeta(2)}{c}\right)^2\eta(y)^{4}\Bigg]\label{hcm:high2}.
\end{align}
In Fig.~\ref{fig:hhcm_dens_high}, using Eqs.~\eqref{hcm:high0}-\eqref{hcm:high2}, we see a good agreement of the analytically obtained series solutions with the numerical solution of Eq.~\eqref{hcm:scaling-chem2_app}.

\section{Nearest neighbor harmonic chain} \label{AppIII}

In this section, we derive the analytical results of the density profiles for the nearest neighbor harmonic chain in a quadratic trap. Recall that, the equilibrium density in the Toda model shows a distinctly different behavior when compared with the HR and HC models as shown in Fig.~\ref{fig:Toda} of the main text. We find that, unlike HR and HC models, the spatial spread of the equilibrium density profile is $N$ independent.  We suspect that this peculiarity is rooted in the fact that the inter-particle repulsion is weak in the Toda model, thereby allowing the particle trajectories to cross. To elucidate the effects of weak inter-particle repulsion in the Toda model, we study the harmonic (Hessian) approximation of the Toda interaction $V_T(r_i)$ [Eq.~\eqref{eqn:toda} of the main text], given by
\begin{align}\label{eqn:harm_app}
	V_H(r_i) = \frac{k_2}{2}r_i^2 - k_3r_i,
\end{align}
where $r_i = x_{i+1}-x_i$, $J>0$ (we set $J = 1$), $k_2 = 1/d^2$, and $k_3 = 1/d$ are the interaction strengths and $d$ determines the length scale of the interaction of the Toda model [Eq.~\eqref{eqn:toda} of the main text]. The subscript $H$ in Eq.~\eqref{eqn:harm_app} stands for the harmonic chain.  Since the harmonic chain in quadratic trap is analytically tractable, it provides a transparent way for understanding the behavior of the Toda model in a suitable parameter regime.

In the following section, we find the equilibrium density profile of the harmonic oscillator in the quadratic trap which has the Hamiltonian
\begin{align}\label{harm:ham}
	H =  \frac{k_1}{2} \sum_{i=1}^N x_i^2 + \frac{k_2}{2}\sum_{i=1}^{N-1} (x_{i+1}-x_i)^2 +k_3(x_1-x_N).
\end{align}
Here $k_1$ is the strength of the quadratic trap, $k_2$ is the spring constant and $k_3$ is the magnitude of the effective external force. The average thermal density is given by
\begin{align}\label{dens:prob}
	\rho_H(x, \beta) = \frac{1}{N}\sum_{l=1}^N \langle \delta(x-x_l) \rangle_{\beta}= \frac{1}{N}\sum_{l=1}^N P(x_l),
\end{align}
where the $\langle . \rangle_{\beta}$ is the thermal average in the canonical ensemble and $P(x_l)$ is the marginal distribution of the position of the $l^{\rm th}$ particle. The task then is to find the marginal distribution, which is
\begin{align}\label{harm:pxl}
	P(x_l = x)  = \frac{1}{Z_{H}}\int_{-\infty}^{\infty} dx_1..dx_l..dx_N &~\delta(x_l-x)\times \notag \\ &{\rm exp}\left(-\beta H\big(\{x_i\}\big) \right),
\end{align}
where the partition function is given by
\begin{align}
	Z_{H} =  \int_{-\infty}^{\infty} dx_1..dx_l..dx_N ~{\rm exp}\left(-\beta H\big(\{x_i\}\big) \right).
\end{align}
Using the variable transformation $y_l = \sqrt{\beta k_2}~x_l$,  Eq.~\eqref{harm:pxl} becomes 
\begin{align}\label{harm:pxl1}
	P\Bigg(x_l = \frac{y}{\sqrt{\beta k_2}}\Bigg)  = \frac{\big(\beta k_2\big)^{\frac{1-N}{2}}}{Z_{H}}&\int_{-\infty}^{\infty} dy_1..dy_l..dy_N ~\delta(y_l- y)\notag \times \\&{\rm exp}\left[-\beta H\Bigg(\Big\{\frac{y_i}{\sqrt{\beta k_2}}\Big\}\Bigg) \right],
\end{align}
We note that the delta function in Eq.~\eqref{harm:pxl1} has the following Fourier representation
\begin{align}\label{harm:delta}
	\delta(y_l -y) = \int_{-\infty}^{\infty}~dq\exp\big[-iq(y_l-y)\big].
\end{align}
Also note that the Hamiltonian in Eq.~\eqref{harm:ham} can be recast in the rescaled variables ${\bf y^T} = [y_1..y_l..y_N]$ as
\begin{align}\label{harm:ham1}
	H = \frac{1}{\beta}\Big[\frac{1}{2}{\bf y^T Ay} + {\bf b^Ty}\Big] &= \frac{1}{\beta}\Bigg[\frac{k}{2} \sum_{i=1}^N y_i^2 + \frac{1}{2}\sum_{i=1}^{N-1} (y_{i+1}-y_i)^2 \notag \\& +\gamma(y_1-y_N) +i q y_l\Bigg].
\end{align}
Here 
\begin{align}\label{harm:defk}
	k = \frac{k_1}{k_2},\quad \gamma = \sqrt{\frac{k_3^2 \beta}{k_2}}~~~~\text{and}\quad b_j = \gamma(\delta_{1,j}-\delta_{N,j})+i q \delta_{jl}
\end{align}
is the $j^{\rm th}$ element of ${\bf b}$. 
Since the interactions are nearest neighbour, ${\bf A} = [A_{ij}]$ is a $N \times N$ tridiagonal matrix given by
\begin{equation}\label{matrixA}
	{\bf A}= 
	\begin{bmatrix}
		\upsilon-1    & -1     &  0     &...      &      0&       0\\
		-1     &  \upsilon     & -1     &...      &      0&       0\\
		0     & -1     &  \upsilon     &...      &      0&       0\\
		\vdots & \vdots & \vdots & \ddots  & \vdots&  \vdots\\
		0      & 0      &  0     &...      &      \upsilon&      -1\\
		0      & 0      &  0     &...      &     -1&     \upsilon-1\\
	\end{bmatrix}
	,
\end{equation}
with
\begin{align}\label{harm:upsilon}
	\upsilon = k+2~.
\end{align}
Using Eq.~\eqref{harm:delta} and Eq.~\eqref{harm:ham1} in Eq.~\eqref{harm:pxl1} we get
\begin{align}\label{harm:probxl}
	P\left(x_l = \frac{y}{\sqrt{\beta k_2}}\right)\notag& = \frac{\big(\beta k_2\big)^{\frac{1-N}{2}}}{Z_{H}}\int_{-\infty}^{\infty}~dq~\exp(iqy)\times\\&\int_{-\infty}^{\infty} dy_1..dy_l..dy_N ~{\rm exp}\left(-\frac{1}{2}{\bf y^T Ay} - {\bf b^Ty}\right).
\end{align} 
Eq.~\eqref{harm:probxl} is a multivariate Gaussian integral which gives
\begin{align}\label{harm:probxl1}
	P\left(x_l = \frac{y}{\sqrt{\beta k_2}}\right)= \frac{\big(\beta k_2\big)^{\frac{1-N}{2}}}{Z_{H}}&\frac{(2 \pi)^{\frac{N}{2}}}{\sqrt{A_N}}\int_{-\infty}^{\infty}~dq~\exp(iqy)\times\notag\\& {\rm exp}\left(-\frac{1}{2}{\bf b^T A^{-1}b} \right),
\end{align} 
where $A_N$ is the determinant of matrix ${\bf A}$ of size $N \times N$. It turns out the integral over $q$ in Eq.~\eqref{harm:probxl1} is Gaussian which can be performed to obtain the normalized distribution of $l^{\rm th}$ particle
\begin{align}\label{prob:l}
	P\left(x_l = x\right) =  \frac{1}{\sqrt{2\pi {\rm Var}(x_l)}}{\rm exp}\left(-\frac{\Big(x-\langle x_l\rangle_{\beta}\Big)^2}{2~{\rm Var}(x_l)}\right),
\end{align}
with the mean position given by
\begin{align}\label{harm:meanxll}
	\langle x_l \rangle_{\beta} =  \frac{\gamma}{\sqrt{\beta k_2}} (A^{-1}_{Nl}-A^{-1}_{1l}),
\end{align}
and the variance given by
\begin{align}\label{harm:varxll}
	{\rm Var}(x_l)=\langle x_l^2 \rangle_{\beta}-\langle x_l \rangle_{\beta}^2 = \frac{A_{ll}^{-1}}{\beta k_2}.
\end{align}
\begin{figure*}[htb]
	\centering
	{\includegraphics[scale=0.73]{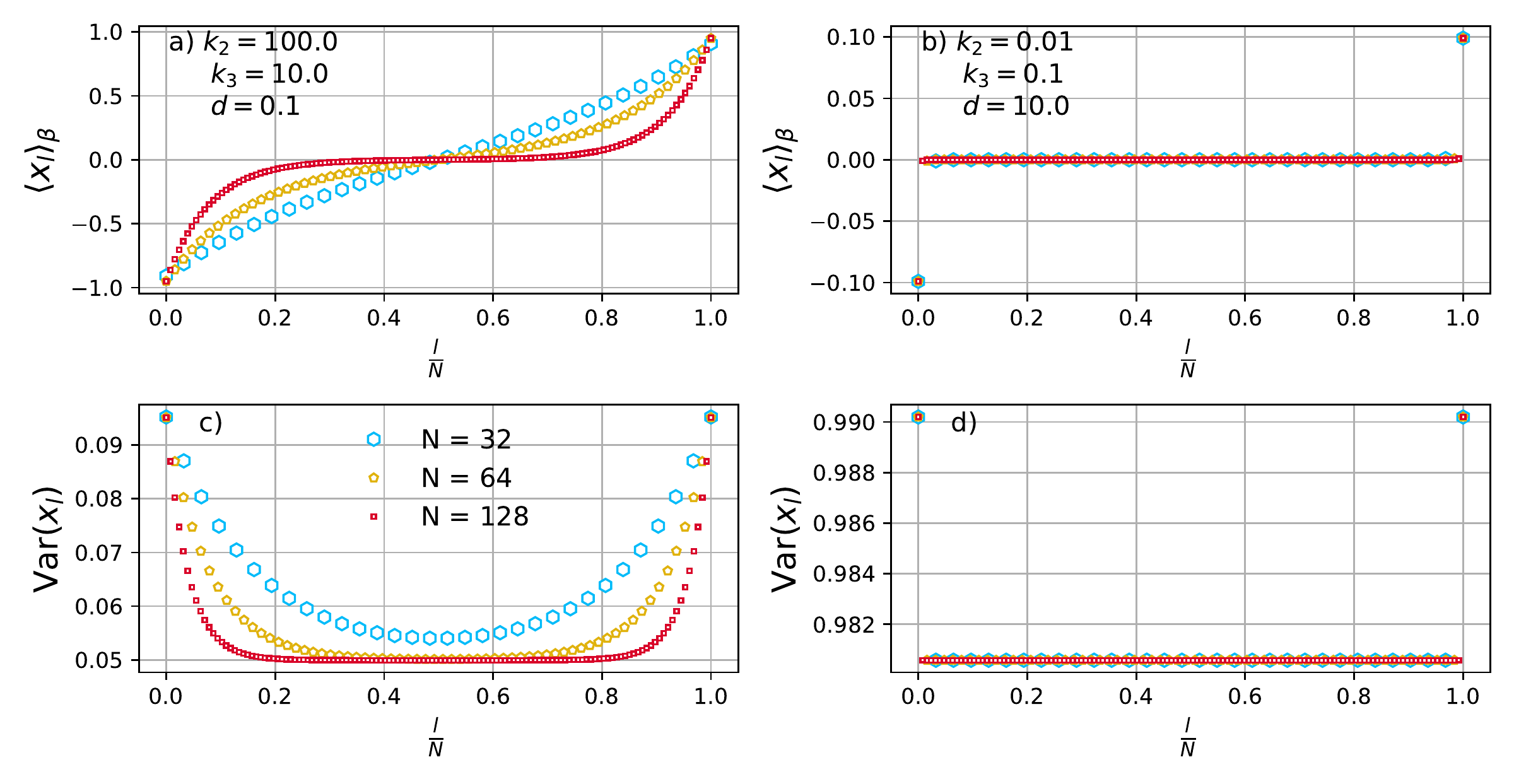}}
	\caption{ (a,b) Mean   and (c,d) variance as function of particle index $l/N$ for the harmonic chain in quadratic trap, computed using Eq.~\eqref{harm:kappal} and Eq.~\eqref{harm:varl} respectively for $N=32, 64, 128$. The $T = 1.0$ and $k_1 = 1.0$. For (a,c) we choose $d = 0.1$ implying $k_2 = 100.0$, $k_3 = 10.0$ and for (b,d) we choose $d = 10.0$ implying $k_2 = 0.01$, $k_3 = 0.1$.}
	\label{fig:harm3}
\end{figure*}
We now have to compute the diagonal elements $A^{-1}_{ll}$, along with the elements in the first and last row of ${\bf A^{-1}}$ i.e., $A^{-1}_{1l}$ and $A^{-1}_{Nl}$. This can be explicitly computed based on the approach in Ref.~\cite{hu1996analytical} which gives
\begin{align}
	A^{-1}_{1l} &= \frac{B_{N-l}}{A_N},\\
	A^{-1}_{Nl} &= \frac{B_{l-1}}{A_N},\\
	A^{-1}_{ll} &= \frac{B_{l-1}B_{N-l}}{A_N}\label{diagAinv},
\end{align}
where recall that $A_N$ is the determinant of the ${\bf A}$ [Eq.~\eqref{matrixA}] and $B_N$ is the determinant of an $N$ dimensional square matrix 
\begin{equation}\label{matrixB}
	{\bf B} = 
	\begin{bmatrix}
		\upsilon    & -1     &  0     &...      &      0&       0\\
		-1     &  \upsilon     & -1     &...      &      0&       0\\
		0     & -1     &  \upsilon     &...      &      0&       0\\
		\vdots & \vdots & \vdots & \ddots  & \vdots&  \vdots\\
		0      & 0      &  0     &...      &      \upsilon&      -1\\
		0      & 0      &  0     &...      &     -1&     \upsilon-1\\
	\end{bmatrix}
	.
\end{equation}
We find that the determinants $A_N$ and $B_N$ are related through the recursion relations
\begin{align}
	A_N &= (\upsilon-1)B_{N-1}-B_{N-2},\label{recA}\\
	B_N &= (\upsilon-1) C_{N-1}-C_{N-2}.\label{recB}
\end{align}
Here $C_N$ is the determinant of an $N \times N$ tridiagonal matrix given by
\begin{equation}\label{matrixC}
	{\bf C} = 
	\begin{bmatrix}
		\upsilon    & -1     &  0     &...      &      0&       0\\
		-1     &  \upsilon     & -1     &...      &      0&       0\\
		0     & -1     &  \upsilon     &...      &      0&       0\\
		\vdots & \vdots & \vdots & \ddots  & \vdots&  \vdots\\
		0      & 0      &  0     &...      &      \upsilon&      -1\\
		0      & 0      &  0     &...      &     -1&     \upsilon\\
	\end{bmatrix},
\end{equation}
where recall that $\upsilon = k+2$.
The determinant of the $N$-dimensional ${\bf C}$ matrix has been computed in Ref.~\cite{hu1996analytical} and is given by
\begin{align}\label{detC} 
	C_N &= \frac{\chi^{-N}}{\chi^2-1}\left(\chi^{2N+2}-1\right),
\end{align}
where 
\begin{align}\label{harm:chi}
	\chi = \frac{\upsilon+\sqrt{\upsilon^2-4}}{2}.
\end{align}
Using the recursion relations Eqs.~\eqref{recA} and~\eqref{recB} with the determinant of $C$ matrix Eq.~\eqref{detC} we get 
\begin{align} 
	B_N &= \frac{\chi^{-N}}{\chi+1}\left(\chi^{2N+1}+1\right),\label{detBN}\\
	A_N &= \frac{\chi^{-N}(\chi-1)}{\chi+1}\left(\chi^{2N}-1\right)\label{detAN}.
\end{align}
Using these expressions, Eqs.~\eqref{detBN} and~\eqref{detAN}, in Eqs.~\eqref{harm:meanxll}-\eqref{diagAinv} we get
\begin{align}\label{harm:kappal}
	\langle x_l \rangle_{\beta} = &\frac{\gamma }{\sqrt{\beta k_2}}\frac{\left(\chi^{N+l}+\chi^{N-l+1}-\chi^{2N-l+1}-\chi^{l}\right)}{(\chi-1)(\chi^{2N}-1)},\\
	{\rm Var}(x_l) \label{harm:varl}&= \frac{\chi}{\beta k_2}\frac{\left(1+\chi^{2N} + \chi^{2N+1-2l} + \chi^{2l-1}\right)}{\left(\chi^2-1\right)\left(\chi^{2N}-1\right)}.
\end{align}
Using the expression of the marginal distribution $P(x_l)$ [Eq.~\eqref{prob:l}] in the expression for average density profile Eq.~\eqref{dens:prob}, we get the expression, for the case of harmonic chain in the quadratic trap
\begin{align}\label{eqn:hrmcdensity}
	\rho_H\left(x, \beta\right) 
	&= \frac{1}{N}\sum_{l=1}^N \frac{1}{\sqrt{2\pi {\rm Var}(x_l)}}{\rm exp}\left(-\frac{\Big(x-\langle x_l\rangle_{\beta}\Big)^2}{2~{\rm Var}(x_l)}\right).
\end{align}
Eq.~\eqref{eqn:hrmcdensity} is the exact result for the harmonic chain in quadratic trap. We now compare Eq.~\eqref{eqn:hrmcdensity} with the Monte-Carlo (MC) densities of the Toda model in quadratic trap in appropriate parameter regimes. 

In Fig.~\ref{fig:harmvstoda}, we show a comparison of the the  MC density profiles of the Toda model, for $N=32, 64, 128$, with the analytically obtained densities of the harmonic chain model [see Eq.~\eqref{eqn:hrmcdensity}], for two values of the Toda interaction length scale $d=1.0, 10.0$. We find a good agreement between the densities of both the models which suggests the $N$-independence of the density profiles. This $N$-independence can be understood in the case of harmonic chain in quadratic trap as follows. For $d \gg 1$, Eq.~\eqref{harm:chi} can be approximated as
\begin{align}
	\chi= \frac{\upsilon+\sqrt{\upsilon^2-4}}{2} \approx d^2, 
\end{align}
where $\upsilon = 2+k = 2+d^2$. Therefore, the mean [Eq.~\eqref{harm:meanxll}] and variance [Eq.~\eqref{harm:varxll}] of the $l^{\rm th}$ particle position are given by
\begin{align}\label{harm:meanldh}
	\langle x_N\rangle_{\beta}= -\langle x_1\rangle_{\beta} \approx \frac{1}{d} &~\text{and}~ \langle   x_l\rangle_{\beta}\approx0~\forall~l \in \{2..,N-1\}\\\label{harm:varldh}
	{\rm Var}(x_l) &\approx \frac{1}{\beta}.
\end{align}
Note that the mean and the variance of the position of the $l^{\rm th}$ particle given in Eqs.~\eqref{harm:meanldh} and~\eqref{harm:varldh} respectively are independent of  $N$ as can be seen from Figs.~\ref{fig:harm3}b and \ref{fig:harm3}d. This proves that the density has no $N$ dependence for large $d$ and therefore Eq.~\eqref{eqn:hrmcdensity} further simplifies to
\begin{align}
	\rho_H(x, \beta) \approx \sqrt{\frac{\beta}{2\pi}}\exp\Bigg(-\beta \frac{x^2}{2}\Bigg).
\end{align}
These $N$-independent density profiles are in accordance with our observations for both Toda and the harmonic chain models [see Figs.~\ref{fig:Toda}c, d and ~\ref{fig:harmvstoda}].

For small values of $d$ the system behaves as a very stiff harmonic chain since the spring constant $k_2 = 1/d^2$ is very large. For these values we expect that the thermodynamic limit can only be attained for extremely large values of $N$. This can be understood more clearly when we consider the asymptotic behavior for the small values of $d$. For small $d \ll 1$, Eq.~\eqref{harm:chi} becomes
\begin{align}\label{harm:chiasymp}
	\chi  = \frac{\upsilon+\sqrt{\upsilon^2-4}}{2}\approx 1+d, 
\end{align}
where $\upsilon = 2+k = 2+d^2$.
Using Eq.~\eqref{harm:chiasymp} in the expressions Eqs.~\eqref{harm:kappal} and~\eqref{harm:varl}, we obtain the mean and the variance of the $l^{\rm th}$ particle position

\begin{align}
	\langle x_l \rangle_{\beta}&= d \left(\frac{1}{(1+d)^{N-l+1}}-\frac{1}{(1+d)^{l}}\right),\\
	{\rm Var}(x_l) &= \frac{d^2}{2\beta}\left(\frac{1}{1+d}+\frac{1}{(1+d)^{2l}}+\frac{1}{(1+d)^{2(N-l+1)}}\right),
\end{align}
for $Nd \gg 1$. In Fig.~\ref{fig:harm3}a and \ref{fig:harm3}c, we show that the mean and the variance, computed using Eqs.~\eqref{harm:kappal} and Eq.~\eqref{harm:varl} respectively, converges very slowly with increasing $N$ to zero and $d/(2\beta)$ respectively. Therefore, we expect the density profiles in Eq.~\eqref{eqn:hrmcdensity} to slowly converge to an $N$-independent profile for very large $N$. The density profile in the large $N$ limit is then given by
\begin{align}\label{eqn:smalld}
	\rho_H(x, \beta) \approx \sqrt{\frac{\beta}{d\pi}}\exp\Bigg(-\beta \frac{x^2}{d}\Bigg),
\end{align}
which is the curve corresponding to $N \to \infty$ in Fig.~\ref{fig:harm2}. This slow convergence is similar to what we find for the Toda model for $d = 0.1$ as shown in Figs.~\ref{fig:Toda}b.
\vspace{1cm}

\bibliography{References}
\end{document}